\documentclass[12pt]{article}

\usepackage{amsfonts}
\usepackage{a4wide}
\usepackage{epsfig}
\newfont{\cyr}{wncyr10}
\newcommand{\news}{\setcounter{equation}{0}}
\newcommand{\be}{\begin{equation}}
\newcommand{\ee}{\end{equation}}
\newcommand{\bea}{\begin{eqnarray}}
\newcommand{\eea}{\end{eqnarray}}
\newcommand{\bean}{\begin{eqnarray*}}
\newcommand{\eean}{\end{eqnarray*}}
\font\upright=cmu10 scaled\magstep1
\font\sans=cmss12
\newcommand{\ssf}{\sans}
\newcommand{\stroke}{\vrule height8pt width0.4pt depth-0.1pt}
\newcommand{\Z}{\hbox{\upright\rlap{\ssf Z}\kern 2.7pt {\ssf Z}}}

\newcommand{\C}{{\rlap{\rlap{C}\kern 3.8pt\stroke}\phantom{C}}}
\newcommand{\R}{\hbox{\upright\rlap{I}\kern 1.7pt R}}
\newcommand{\CP}{\C{\upright\rlap{I}\kern 1.5pt P}}
\newcommand{\PP}{\hbox{\upright\rlap{I}\kern 1.5pt P}}
\newcommand{\taubf}{\mbox{\boldmath $\tau$}}

\input mssymb

\begin{document}
\pagestyle{plain}

\title{\vskip -70pt
\begin{flushright}
{\normalsize DAMTP  98--26} \\
{\normalsize hep-th/9804142}
\end{flushright}
\vskip 20pt
{Zero mode quantization of multi-Skyrmions} \vskip 10pt
}

\author{Patrick Irwin\thanks{
Email. pwi20@damtp.cam.ac.uk 
} \\[10pt]
{\sl Department of Applied Mathematics and Theoretical Physics} \\[5pt]
{\sl University of Cambridge} \\[5pt]
{\sl Silver St., Cambridge CB3 9EW, England}\\[5pt]
 \\[10pt]}

\date{August 1998}
\maketitle
\begin{abstract}
A zero mode quantization of the minimal energy SU(2) Skyrmions for nucleon
numbers four to nine and seventeen is described. This involves quantizing the
rotational and isorotational modes of the configurations. For nucleon 
numbers four, six and eight the ground states obtained are in
agreement with the observed nuclear states of Helium, Lithium and Beryllium. 
However, for nucleon numbers five, seven,
nine and seventeen the spins obtained conflict with the observed
isodoublet nuclear states.  

\end{abstract}

\newpage 
\section{Introduction}
\news
\ \indent 
In this article a simple quantization of higher charge Skyrmions is
described and the results are compared to experimental nuclear
data. The methods described may be used for Skyrmions
of any nucleon number $B$, once the minimal energy solution
is known. The minimal energy solutions are now known for $B\leq 9$ 
\cite{BS} and a conjectured solution exists for $B=17$  \cite{HMS}.
We use the moduli space
approximation \cite{NSM}, which truncates the infinite dimensional
configuration field space to a finite dimensional space consisting
of classical configurations which are relevant to the low energy
dynamics. The moduli space will necessarily include all minimal energy
configurations and to obtain accurate results one should include all
configurations corresponding to $B$ Skyrmions with arbitrary separations
and relative isospin orientations. Obviously, the more configurations
that are included in the moduli space, the more difficult  
their analysis becomes. As a first approximation one may restrict the
moduli space to be generated by the zero modes of the minimal energy solution.
Any Skyrmion configuration can be translated, rotated or isorotated 
without changing its energy and these are the only zero modes. We
shall ignore the translational modes since their quantization only
gives a total momentum to the quantum state. The interesting physics
arises when the isospin and spin degrees of freedom are quantized.

The minimal energy Skyrmions for
$B=1$ and $B=2$ have spherical and axial symmetry respectively. For higher
nucleon numbers the minimal energy solutions only have a discrete 
symmetry \cite{BS, BCT}. This means that the classical configuration 
$U_B({\bf x})$ is invariant under a discrete group, $H$, 
of combined rotations and 
isorotations. 
Thus the moduli space of zero modes 
is given by a quotient space ${\cal C}=(\mbox{SO(3)}
\times\mbox{SO(3)})/H$. 
This may be equivalently written as a quotient
of the covering group 
${\cal C}=(\mbox{SU(2)}\times\mbox{SU(2)})/K$, 
where $K$ is a discrete subgroup of SU(2)$\times\,$SU(2) related to
the discrete subgroup $H$ of SO(3). Elements of $K$ correspond to
rotations and isorotations in $H$ combined with $2\pi$ rotations
and isorotations. In the cases where we need to be specifically 
concerned with $K$, as opposed to $H$,
it has the form $K=\bar{H}\times\Z_2$ where $\bar{H}$ is
the double group of $H$ ($\bar{H}$ is a subgroup of
SU(2) with two elements, $h$ and $-h$ in $\bar{H}$ for every element
of $H$ in SO(3)). 
Elements in $K$ distinguish a clockwise rotation by $\theta$ 
about some axis from an anticlockwise rotation by
$2\pi-\theta$ about the same axis. In the even
nucleon sector $2\pi$ rotations are trivial, so it is sufficient to
consider the group $H$. But for odd $B$ it is necessary to consider
$K$ as opposed to $H$.

Semiclassical quantization of the configuration is achieved by
quantizing on this quotient space. There are a
number of inequivalent ways to quantize on a quotient space $G/K$;
when $G$, which here is $\mbox{SU}(2)\times\mbox{SU}(2)$, is simply 
connected these are labelled by the irreducible 
representations of the group $K$.
In general, the wave functions are defined on SU(2)$\times$SU(2), 
but they transform under some irreducible representation of $K$.
The reason for working with the double cover, SU(2)$\times$SU(2),
is that as is
well known, $2\pi$ rotations have nontrivial consequences in the quantum
theory, this enables single Skyrmions to be quantized as fermions. 
To determine which quantization is appropriate here, one must consider the
Finkelstein-Rubenstein (FR) constraints \cite{FRW}.  
They state that, in order for a single Skyrmion to be quantized as
a fermion, wave functionals are sections of a 
line bundle over the classical configuration space whose holonomy
around any noncontractible loop in the configuration space is (-1). 
In our case, quantizing on ${\cal C}$, wave functions are sections of a
line bundle over ${\cal C}$ whose holonomy is (-1) for loops which remain
noncontractible when ${\cal C}$ is extended to the 
full Skyrmion field configuration space. This is equivalent 
to defining wave functions  
on SU(2)$\times$SU(2) which are eigenstates of the operators which
correspond to a rotation and isorotation by an element of $K$, 
with eigenvalues (-1) +1 depending on
whether this operation is (non)contractible in the full Skyrmion
configuration space. The effect of 
$2\pi$ rotations or isorotations is well known. A $2\pi$
rotation or isorotation of a
configuration with nucleon number $B$ is contractible if $B$ is even
and noncontractible if $B$ is odd. Thus states with odd $B$
are fermionic and states with even $B$ are bosonic. 
The states define a one dimensional
representation of the symmetry group $K$. If $K$ has 
no nontrivial one
dimensional representations then all the FR constraints must be +1. If
there are nontrivial one dimensional representations 
of $K$ then one needs to carefully
examine the closed loop corresponding to elements of $K$ which have
character (-1) of this nontrivial one dimensional representation.
It must be determined whether these loops are contractible or not.

For this, it is often necessary to split the configuration 
into individual or pairs of well separated Skyrmions, 
and then analyse the closed loop. The (non)contractibility of the $B=1$
and $B=2$ Skyrmions are known under such
closed loops and from this the (non)contractibility of the loop may be
determined. However, it is necessary that the configuration retains
the symmetry of the specific element of $K$ being considered, as 
it is being split into a well separated configuration of Skyrmions, 
i.e. the loop is closed throughout the deformation.
This is not obvious from the Skyrme picture since there is no
analytical data. 

To proceed we can use 
the recently discovered rational map
ansatz for Skyrmions \cite{HMS}. These authors describe
how, given an SU(2) monopole which can be uniquely described by a 
rational map, one may
associate to it a Skyrmion. Using this method they were able to
accurately approximate the known minimal energy Skyrmion
configurations for nucleon numbers one to nine and the predicted
solution for nucleon number seventeen. The minimal energy
Skyrme configuration obtained in this manner has the same symmetries
as the monopole from which it is derived. This ansatz has the
advantage of clearly illustrating what combination of rotations and
isorotations leave the solution invariant. It is also useful in that the
reflection symmetries of the solution can easily be worked out which 
enables one to determine how the parity operator can be represented on
${\cal C}$. 

As verified in \cite{BBT} this ansatz 
also extends to describe some of
the Skyrmions vibrational modes. There, the vibrational
spectra of the minimal energy $B=2$ and $B=4$ Skyrmions was calculated. 
The vibrations form representations of the symmetry group of the
minimal energy Skyrmion. The vibrational modes of the Skyrmions
come in two different types. The modes of lower frequency 
correspond to the Skyrmion configuration breaking up into separated
Skyrmions. The modes of higher frequencies correspond to the well known
``breather'' and generalisations of it whereby the local nucleon charge
expands or contracts in places (in \cite{BM} a mechanism was given for
describing these modes). 
It is also possible to look at vibrations of the rational maps. This
corresponds to monopole motion on the monopole moduli space. Again, small
variations from the symmetric configuration form representations 
of its symmetry group.
In \cite{BBT}, they found that 
vibrations with frequency below that of the ``breather'' type modes
form the same representations of the symmetry group as do the monopole
vibrations. The implication of this is that, if a monopole
configuration can be separated a small distance while respecting some 
discrete symmetry, then the same process can occur for Skyrmions. 
We wish to extend this
correspondence to arbitrary monopole/Skyrmion separations.
However, the rational map ansatz breaks down as the monopole separates
into individual Skyrmions. Nonetheless we conjecture that the
correspondence can be extended beyond this region, such that any
monopole motion can be mapped to an equivalent path in the Skyrmion
configuration space. In effect, this amounts to an embedding of the
monopole moduli space into the Skyrmion configuration space.
Evidence for this is seen by considering the possible scattering
processes for the known cases of monopoles and Skyrmions 
\cite{SD}. In the above paper it was seen that 
for well known cases of monopole scattering, an equivalent 
Skyrmion scattering process could occur with the same symmetry.
In fact, all we really need to assume is that, if the Skyrmion can be vibrated
remaining invariant under some symmetry group element, then the
continuation of this path in the Skyrmion configuration space, which will
remain invariant under the symmetry, eventually becomes a
configuration of well separated Skyrmions. For the monopoles this is
always the case.   

Assuming the results in \cite{BBT} are true for general nucleon
numbers, and that there is a 1-1
correspondence between monopole motion and Skyrmion motion for low
vibrational energies then, if the monopole configuration can be deformed
keeping a symmetry, so can the Skyrmions. 
But monopoles are in an exact 1-1 correspondence with  
rational maps \cite{J, D}. 
The set of monopoles which have a discrete rotational symmetry is
easily determined from the rational maps (because they have a simple
action of the rotation group). Also, it is easy to see how the
rational map of a symmetric multi-monopole changes when the 
multi-monopole splits up into well separated monopoles. So, rational maps
can be used to determine whether a multi-monopole can be split into a
specific configuration of well separated monopoles while respecting a
certain symmetry group element. Thus by our above assumption it can be
determined how a Skyrmion configuration can be split up while keeping
a certain symmetry. In the cases considered here we can always
separate into a configuration of $B=1$ and $B=2$ solutions whose
behaviour under rotations and isorotations is known. 
Using this method we
shall determine the FR constraints. Once these are found it is a
simple exercise to find the allowed quantum states.

In \cite{BC} and \cite{C}, such an analysis was carried out for
the axially symmetric charge two solution and for the tetrahedrally 
symmetric charge three solution. For the $B=2$ case a ground state
with the correct quantum numbers of the deuteron was obtained. 
And for $B=3$ it was found that
the ground state had spin $\frac{1}{2}$, isospin $\frac{1}{2}$ 
in agreement with the observed isodoublet nucleus 
$(^3_1\mbox{H},\,^3_2\mbox{He})$. 
Here we extend the analysis to the minimal energy Skyrmions with 
nucleon numbers four to nine and seventeen.

We find that for $B=$4, 6, 8 the ground state has the correct 
spin, parity and isospin assignments as for
$^4_2\mbox{He}^+,\,^6_3\mbox{Li}^+$ and
$^8_4\mbox{Be}^+$. However for odd nucleon numbers 
$B=$5, 7, 9 or 17 the ground states
found by this method do not agree with the observed isodoublet states. 
For nucleon numbers 5, 7 and 9 the experimentally observed ground states
are isodoublets with spin $\frac{3}{2}$ and for $B=17$ the observed
ground state is an isodoublet with spin $\frac{5}{2}$ \cite{N}. 
However the zero mode quantization of Skyrmions results in the 
ground state for $B=5$ and $B=9$ both to be isodoublets
with spin $\frac{1}{2}$, for $B=7$ and $B=17$ the ground state are both 
found to be isodoublets with spin $\frac{7}{2}$. As discussed below we
do not try to predict the parity of the states with odd $B$. 
The ground states we find here exist experimentally as excited states.
The experimentally observed
ground state for $B=5$ appears here as an excited state.
The experimentally observed ground states for $B=7$, $B=9$ and
$B=17$ can be obtained here by including the vibrational modes but
they will also appear here as excited states.

The vibrational modes form representations of the symmetry group 
of the minimal energy solution. Knowing this it is possible to
combine the rotational and vibrational modes resulting in an enlarged
configuration space. The vibrational spectra has been worked out for
the Skyrme model for nucleon numbers two \cite{BBT, WAL} and four
\cite{BBT}, it is also possible to understand some 
aspects of the vibrational spectra
for other values of $B$ using the rational map ansatz.
The vibrational modes of the Skyrmions with frequencies below the 
breather modes can be described by monopole motion and thus the 
representations they form of the symmetry group can be determined.
The configuration space is now a fiber bundle
over $(\mbox{SU}(2)\times\mbox{SU}(2))/K$, the fiber 
being the vector space
corresponding to the vibrations. This space
was described in \cite{GK}. States are now given by the direct product
of Wigner functions on SU(2)$\times$SU(2) and harmonic oscillator
wavefunctions on the vibrational space. The states must satisfy a $K$
invariance condition described below which restricts the allowed set
of states. Using this formalism further excited states of the multi-Skyrmions
may be described. It is possible that this approach may resolve the
above problem of the ground state for $B$=7. A spin $\frac{3}{2}$ 
rotational state can be combined with a vibrational state to give an
allowed state. If the vibrational energy of this state is not too
large it may  have lower energy than the state with spin
$\frac{7}{2}$ and thus predict the correct ground state. To check
this, the energies of the vibrational states need to worked out directly from
the Skyrme model as the rational map approach has no information about
the frequencies of the specific vibrations. 
The inclusion of vibrational modes may also fix the problem for $B=17$
but it will not work for $B=5$ and $B=9$.

Naturally, one would not expect that the quantization of
zero modes and vibrations would give accurate results 
on binding energies of the
states etc., and the inclusion of more degrees of freedom are needed to
accurately describe such properties. Nonetheless it is
not obvious that the inclusion of other modes (allowing the
Skyrmions to separate, calculating the zero point energies of the
radiative pion modes) will resolve this difficulty. A possible
resolution of this is that the solutions found in \cite{BS} are not
well defined minima, i.e. there may be a number of local minima with 
approximately equal energies and so an expansion about just one of
these minima is not valid. This seems to occur for the $B=10$ case,
to answer the question here requires further numerical investigation
of the proposed minimal energy solutions.  

In the following section
we review the zero mode quantization discussed in \cite{BC} paying
special attention to the FR constraints. Section 3 describes
the rational map ansatz for Skyrmions and how it may be used to
determine the FR constraints. Then in section 4 the 
quantization procedure is treated for each of the Skyrmions $B=4$ 
to $B=9$ and $B=17$. Section 5 describes how to include vibrational
modes and gives the predicted excited states for the $B$=4 sector by
considering the vibrations together with the zero modes.
Finally in section 6 we calculate the expectation value of the nucleon 
density of the quantum ground states and compare to the 
classical nucleon densities. A criticism raised about the classical
solutions of the Skyrme model is that they bear no resemblance to real
nuclei. The classical nucleon densities have the symmetry of some
discrete group. To find the nucleon density in the quantum state,
following \cite{LMS} we integrate the classical nucleon density times
the norm squared of the wave function over the
moduli space. We find that in all cases considered, the nucleon density
in the quantum state is almost spherically symmetric, being exactly so
in a number of cases. For example we find the ground state for $B=4$
to be spherically symmetric and for $B=6$ to be mainly S-wave with a
small P-wave admixture. This agrees with the nucleon densities
of Helium 4 and Lithium 6 respectively and shows how the nucleon
density of the classical solutions is smeared by quantum effects to
a more uniform angular dependence.

\section{Semi-Classical Quantization}
\news

The Skyrme model has the Lagrangian
\be L = \int d^3x\left\{ -\frac{f^2_{\pi}}{16}\mbox{Tr}({R^{\mu}}{R_{\mu}})
+\frac{1}{32e^2}\mbox{Tr}([R_{\mu} ,R_{\nu} ][R^{\mu} ,R^{\nu}])
\right\}\label{1}
\ee
where $R_{\mu}=\partial_{\mu}UU^\dagger$, $\,U$ is the SU(2) valued 
Skyrme field, and $e$, $f_{\pi}$ 
are free parameters of the model whose values are chosen to
best fit experimental data.
The above Lagrangian has soliton solutions of finite energy. 
Finite energy implies that $U$ tends to a constant 
at spatial infinity. Space is then
compactified to $S^3$ and thus each soliton solution has an associated
integer, the degree, corresponding to the element of
$\pi_{3} (S^3)$ to which $U$ belongs. Solitons of
degree $B$ are interpreted as $B$ nucleons \cite{THR}.

The symmetry group of the Skyrme Lagrangian is 
SO(3)$\times\mbox{  Poincar\'{e} Group}\times \mbox{P.}$ 
P is the parity operator which acts as
P : $U[{\bf x}]\rightarrow U^{\dagger}[-{\bf x}]$.
For time-independent 
fields such as static solitons the symmetry group
is reduced to
\be\mbox{SO(3)}\times \mbox{Euclidean Group of } 
\R^3\times \mbox{P}. 
\ee
The minimal energy solutions to the Skyrme model, $U_B[{\bf x}]$,
have for $B\geq 3$ a
discrete symmetry group.
This means that the classical configuration $U_B[{\bf x}]$
is invariant under a discrete group $H$, of combined rotations and
isorotations. 
To every element $S\in\,H$ there will exist an element
$\Gamma(S)\in\,$SO(3) such that the rotation $S$, 
has the same effect on the configuration as 
the isorotation $\Gamma(S)$. Or alternatively, the combined rotation
$S$ and isorotation $\Gamma^{-1}(S)$ leaves the configuration unchanged. 
The elements $\Gamma(S)$ form a representation
of the group $H$. This is true because to each rotation $S$,
$\Gamma(S)$ is unique. If $\Gamma(S)$ was not unique then    
the Skyrmion would be invariant under an isorotation, without any
compensating rotation. Assuming this isorotation is 
about the $x_3$-axis,
a simple argument shows that the right currents $R_i$ are proportional
to $\tau_3$ (with $\tau_i$ the Pauli matrices).
But this implies that the nucleon density ${\cal B}$ must vanish
because it is given by 
\be\label{bdensity}
{\cal B}=\frac{1}{24\pi^2}{\epsilon_{ijk}
\mbox{Tr}\,R_{i}R_{j}R_{k}}\;.
\ee
For $B=3$ and $B=9$ $U_B$ has tetrahedral
symmetry, $B=4$ and $B=7$ have octahedral and icosahedral symmetry
respectively and the $B=5$, $B=6$ and $B=8$ solutions 
have $D_{2d}$, $D_{4d}$, and $D_{6d}$ symmetries respectively.

One can act on the classical solutions with 
$\R^3\times\,$SO(3)$\times$SO(3) 
in the following fashion to give a family of solutions with the same 
energy. This generates the zero mode moduli space.
The transformations correspond to translations, rotations and isospin 
rotations. Explicitly
\be 
U_B[{\bf x}]\rightarrow
 A'U_B [D(A)({\bf x}-{\bf a})]A'^\dagger\label{2.3}
\ee 
where $A,\,A'$ are in the fundamental representation of SU(2),
${\bf a}\in \R^3$, and 
$D(A)$ is the SO(3) element associated to $A$, given by
$D(A)_{ij}=\frac{1}{2}\mbox{Tr}\,\tau_iA\tau_jA^{-1}$.
So generally the minimal energy solution will have 
nine zero modes. We will henceforth ignore the translational 
$\R^3$ symmetry. 
The above is an SO(3)$\times$SO(3) action since $A$ 
has the same effect on $U_B[{\bf x}]$ as $-A$ does, and similarly for
$A'$. 
If we label an element in the moduli space by $\{A,\,A'\}$ we
have the identifications
\be\label{2.223}
\{A,\,A'\}\cong \{A,\,-A'\}\cong \{-A,\,A'\}\cong\{-A,\,-A'\}\;.
\ee
The moduli space approximation to multi-Skyrmion dynamics  involves
letting $A,\,A'$  become time-dependent and substituting (\ref{2.3})
into the Skyrme Lagrangian (\ref{1}). 
The reduced Lagrangian is quadratic in the time
derivatives $a_k=-i\mbox{Tr}\,\tau_kA^{'\dagger}\dot{A}^{'},\;
b_k=-i\mbox{Tr}\,\tau_kA\dot{A}^{\dagger}\!\!\!$, and is given by
\be\label{2.5}
L_B=\frac{1}{2}a_iU_{ij}a_j+\frac{1}{2}b_iV_{ij}b_j-a_iW_{ij}b_j-M_B
\ee 
with $M_{B}$ is the mass of the
solution and 
the tensors $U_{ij},V_{ij},W_{ij}$ are dependent on
the classical solution $U[{\bf x}]$, given by \cite{BC}
\bea\label{2.6}
U_{ij}&=&\frac{1}{8}\int d^3x\mbox{Tr}\left
\{U^{\dagger}[\frac{1}{2}
\tau_{i},U]U^{\dagger}[\frac{1}{2}\tau_{j},U]+[U^{\dagger}\partial_kU,
U^{\dagger}[\frac{1}{2}\tau_{i},U]][U^{\dagger}\partial_kU,
U^{\dagger}[\frac{1}{2}\tau_{j},U]]\right\}\nonumber\\
W_{ij}\!\!&=&\!\!\frac{i}{8}\int d^3x\mbox{Tr}\left\{
U^{\dagger}[\frac{1}{2}\tau_{i},U]U^{\dagger}
({\bf x}\times\nabla)_jU+[U^{\dagger}\partial_kU,
U^{\dagger}[\frac{1}{2}\tau_{i},U]][U^{\dagger}\partial_kU,
U^{\dagger}({\bf x}\times\nabla)_jU]\right\}\!\!\!\!\nonumber\\
V_{ij}&=&-\frac{1}{8}\int d^3x\mbox{Tr}\left\{U^{\dagger}
({\bf x}\times\nabla)_iUU^{\dagger}({\bf x}\times\nabla)_jU\right\}
\nonumber\\&-&\frac{1}{8}\int d^3x\mbox{Tr}\left\{
[U^{\dagger}\partial_kU,
U^{\dagger}({\bf x}\times\nabla)_iU][U^{\dagger}\partial_kU,
U^{\dagger}({\bf x}\times\nabla)_jU]\right\}
\eea
where we have set $f_{\pi}^2=8$ and $e^2=1/2$. 

This Lagrangian may now be quantized in
the manner described in \cite{BC}. 
The momenta conjugate to $a_i$ and
$b_i$ become the body-fixed
spin and isospin angular momentum operators called $K_i$ and
$L_i$ which satisfy the SU(2) commutation relations, 
$[K_i,K_j]=i\epsilon_{ijk}K_k$, and similarly for $L_i$.
There also exist space-fixed spin and isospin angular momentum
operators denoted by $J_i$ and $I_i$ related to the body-fixed
operators by 
\be
J_i=-D_{ij}(A)^TL_j\,,\;\;I_i=-D_{ij}(A')K_j\,.
\ee
The commutation relations are
\be\label{cccr}
[L_i,A]=-\frac{1}{2}\tau_iA\,,\;\;\;\;\;[J_i,A]=
\frac{1}{2}A\,\tau_i\,,\;\;\;\;\;
[I_i,A']=-\frac{1}{2}\tau_iA'\,,\;\;\;\;\;[K_i,A']=
\frac{1}{2}A'\tau_i
\ee
and all other commutators vanish.
This means that ${\bf L}^2={\bf J}^2$ and ${\bf I}^2={\bf K}^2\,.$
The Hamiltonian becomes that of a rigid body in space and isospace.
However, the above derivation of the rigid body Hamiltonian 
is not complete since we have not considered the
discrete symmetry group $H\subset\,$SO(3), of the solution.
This means that rotating the configuration by an element
$S\in\,$SO(3) has the same effect on the configuration as 
the isorotation $\Gamma(S)$.
The isorotations
need not be the same as the rotations, but they do form
a three dimensional representation of $H$. 
Labelling $\{R,R'\}$ as the set of zero modes corresponding to
rotations and isorotations,
SO(3)$\times$SO(3), we have
the following identification 
\be\{R,R'\}\cong\{SR,\,R'\Gamma^{-1}(S)\}\;,\;\;S\in\,H\,.
\ee
Thus the moduli space is $(\mbox{SO}(3)\times 
\mbox{SO}(3))/H$ with the above quotient. 
But we really need to consider the covering space 
SU(2)$\times$SU(2) because $2\pi$ rotations or
isorotations can be noncontractible. 
If we view the moduli space as a quotient of 
SU(2)$\times$SU(2) then each closed loop corresponding to
$\{S,\,\Gamma(S)\}$ will correspond to four closed loops 
since both $S$ and $\Gamma(S)$ 
can be lifted in two ways to SU(2).
We now have the identifications
\be\label{2.222}
\{A,A'\}\cong \{hA,\,A'h'^{-1}\}\;,\;\;h\in\,\bar{H}\,,
\ee
where $h$ and $h'$ are in the fundamental representation of SU(2) 
and $D(h)=S$ and $D(h')=\Gamma(S)$. 
So $\pm h$ and $\pm h'$ are lifts to SU(2) 
of $S$ and $\Gamma(S)$ respectively. 
The elements $h$ form the double group $\bar{H}$ of $H$.
(\ref{2.222}) includes (\ref{2.223})
and determines the moduli space as 
SU(2)$\times$SU(2)$/K$ where $K$ is the subgroup of SU(2)$\times$SU(2)
consisting of the elements $\{\pm h,\,\pm h'\}$.
If the representation $\Gamma$ of $H$ lifts to a representation
$\tilde{\Gamma}$ of $\bar{H}$ i.e. such that
$D(\tilde{\Gamma}(h))=\Gamma(D(h))$  for $h\in\bar{H}$,  
with $\tilde{\Gamma}(h_1h_2)=\tilde{\Gamma}(h_1)\tilde{\Gamma}(h_2)$ and 
$\tilde{\Gamma}(-1)=-1$, then $K$ has the form
$\bar{H}\times \Z_2$, but it is not always possible that $\Gamma$ can be
lifted in this way.

Each element of the group $K$ corresponds to one of
the four ways of lifting a symmetry group element in $H$ to the
covering space. For each element of $K$
it is necessary to determine whether this
transformation is a contractible loop in the Skyrmion configuration
space or not. When $B$ is even this task is simpler since $2\pi$
rotations and isorotations are contractible so there is no need to
distinguish a rotation from the same rotation plus a $2\pi$
rotation. But in the odd $B$ case this distinction matters and so it
is necessary to be more careful. We deal with this on a case by
case basis in section 4. 

As discussed in the introduction, to determine which quantization is
appropriate here we need to consider the 
Finkelstein-Rubenstein (FR) constraints \cite{FRW}.
These authors showed that it is possible to quantize the solitons as
fermions if one lifts the classical configuration space to its
simply connected covering space. A quantization scheme which
treats single Skyrmions as
fermions is to multiply states by a phase +1 (-1) when acted on by
operators which implement contractible (noncontractible) loops in the
classical configuration space.
They also showed that the exchange of
two identical Skyrmions and the $2\pi$ rotation of one of the 
Skyrmions are homotopic loops thus proving that the usual notion of 
spin-statistics holds in the Skyrme model.
Also, as a result of the fact that $\pi_4$(SU(2))=$\Z_2$ there are
only two topologically distinct loops in the space. Williams
\cite{W} verified that the $B=1$ Skyrmion can be quantized as a fermion by 
showing that a $2\pi$ rotation of it is a
noncontractible loop in the Skyrmion configuration space. This was
extended in \cite{G} whereby it was shown that the $2\pi$ rotation of
a charge $B$ Skyrmion is contractible if $B$ is even and
noncontractible is $B$ is odd. 

Thus the operator which corresponds to
implementing a closed loop on the configuration space acts on states
with eigenvalue $\pm 1$ according to the contractibility of the
loop. In our case the closed loops always correspond to rotations or
isorotations and the operators which generate such transformations are
${\bf L}_i$ and ${\bf K}_i$.
If the symmetry group element is of the form
\be\label{2.141}
\{h,\,h'\}=\{e^{-\frac{i\theta_1}{2}\hat{{\bf n}}_1\cdot{\bf \taubf}},\,
e^{-\frac{i\theta_2}{2}\hat{{\bf n}}_2\cdot{\bf \taubf}}\}\,,
\ee
then using (\ref{cccr}) and (\ref{2.222})
the constraints on the quantum states arising from the symmetries  of 
the classical solution may be expressed as
\be\label{2.14}
e^{i\theta_1\hat{{\bf n}}_1\cdot{\bf L}}
e^{i\theta_2\hat{{\bf n}}_2\cdot{\bf K}}|\Psi>=\pm|\Psi>\,,
\ee
the $\pm$ depending on whether the loop corresponding to
$\{h,\,h'\}$ is contractible or not in the full configuration 
space. A $2\pi$ rotation or isorotation of a Skyrmion of 
nucleon number $B$ is contractible if $B$ is even and
noncontractible if $B$ is odd. So, physical states $|\Psi>$ also satisfy 
\be\label{2.15}
e^{2\pi i\hat{{\bf n}}\cdot{\bf K}}|\Psi>=e^{2\pi i\hat{{\bf n}}\cdot{\bf
L}}|\Psi>=(-1)^B|\Psi>
\ee
This means that for even $B$, $I$ and $J$ are
integral and for odd $B$, $I$ and $J$ are half integral.

Returning to the Lagrangian in (\ref{2.5}), in general 
$U_{ij},\,V_{ij},\,W_{ij}$ are diagonal. 
The number of different eigenvalues of $U_{ij},\,V_{ij},
\,W_{ij}$
depends on the symmetry of the solution. Tetrahedral, octahedral or
icosahedral symmetry implies the matrices have a single eigenvalue if
the fields transform according to a three dimensional irreducible 
representation of the group. For instance, the 
$B=4$ solution has octahedral symmetry 
whereby a rotation by an element of the octahedral group combined with an
isorotation leaves the solution invariant. The rotations form the
defining representation of the octahedral group and so $V_{ij}$ is
proportional to the identity matrix with one common moment of
inertia. But the corresponding isospin transformations are in a
reducible representation, comprising irreducible representations of
dimensions one and two and this means that $U_{ij}$ has two 
distinct eigenvalues.$\footnote{I thank K. Baskerville for pointing 
this out to me.}$ 
It also turns out that the cross term $W_{ij}$ vanishes because the
symmetry is realised differently between the spin and isospin.

If the matrices have only one eigenvalue the Hamiltonian is that of a
spherical top \cite{LL} (in space and isospace). If the Skyrmion 
has an axis of symmetry above the second order (for $B=6$ and $B=8$) 
then $U_{ij},\,V_{ij},\,W_{ij}$ have two distinct 
eigenvalues and the Hamiltonian is that of a symmetrical top (in such
cases we take the axis of symmetry to be the $x_3$-axis),
otherwise (for $B=5$) $U_{ij},\,V_{ij},\,W_{ij}$ has 
three different eigenvalues and
the Hamiltonian is that of an asymmetrical top. A basis for the 
Hilbert space of states is given by 
$|J,J_3,L_3>\otimes\;|I,I_3,K_3>$, with 
$-J\leq J_3,\,L_3\leq J$ and $-I\leq I_3,\,K_3\leq I$. In all that follows, 
the third component of the space and isospace
angular momentum $J_3$ and $I_3$ are omitted. 
The value of $J_3$ corresponds to the angular momentum eigenvalue of
the state about a fixed axis in space and is not physically relevent.
States with differing values of $I_3$ correspond
to the different states in an isospin multiplet, e.g. $I_3=2$ means the
state has two more protons than neutrons etc. These states
will be energy eigenstates in all cases except possibly for $B=5$, 
where the energy eigenstates will not generally have 
definite values of $K_3,\,L_3$. 

It is an easy numerical task to calculate the moments of inertia from
the rational map generated Skyrmions. As described in the next section
the Skyrme field is approximated by 
$U[r,\theta,\phi]=\exp(if(r)\hat{{\bf n}}_R\cdot\taubf)$ 
where $\hat{{\bf n}}_R$
is derived from a rational function of
$z=\tan(\theta/ 2)e^{i\phi}$, where $r$, $\theta$ and $\phi$ are polar
coordinates and $f(r)$ is determined numerically. Inserting this into
(\ref{2.6}) the moments of inertia are obtained by radial and angular
integrations. It is found that the rotational moments of inertia
$(V_{ij})$ become much larger than the isorotational moments of
inertia $(U_{ij})$ as $B$ increases. For example, 
for $B=1$ the moment of inertia is 
$U_{ij}=V_{ij}=106.4\delta_{ij}$ in units of $1/e^3f_{\pi}$, the
rotational and isorotational moments of inertia being equal due to
spherical symmetry. But already at $B=4$ we have
$U_{11}=U_{22}=254.0\,,\;U_{33}=306.4$ 
and $V_{11}=V_{22}=V_{33}=1162.9$. $U_{ij}$ 
and $W_{ij}$ increase approximately like $B$
while $V_{ij}$ increases like $B^2$ (of course in certain cases some
symmetry can imply that some moments of inertia are zero).
The energies of rotational states are like $\frac{1}{2}J(J+1)/V$, and
isorotational states are like $\frac{1}{2}I(I+1)/U$ 
where $V$ and $U$ indicate
the rotational and isorotational moments of inertia 
and $J$ and $I$ indicate the spin 
and isospin eigenvalues. We see that states
with the lowest energy will always have $I$ as small as
possible (there
is a contribution from the $W$ moments of inertia but since these are
of order $U$ it doesn't change the outcome). So states with high
isospin are energetically unfavourable and will not exist, this is
true of real nuclei whose nucleon number is small. As the nucleon
number increases, electromagnetic effects will favour neutrons over protons
but for all small nuclei ($B\leq 30$) the ground state has the
smallest possible value of isospin. 
So to find the lowest energy states we will always set the isospin to
its lowest possible value.

To obtain the correct quantum states we need to determine the
(non)contractibility of the closed loops, corresponding to elements of
$K$, in the configuration space.
To do this we use the rational map description of Skyrmions which we
now review.

\section{Rational Map Generated Skyrmions}
\news

To describe the symmetries of the Skyrmions, and thus 
evaluate the FR constraints, we shall use the rational map ansatz 
for Skyrmions which was introduced in \cite{HMS}. 
Jarvis has shown that there is a 1-1 correspondence
between SU(2) monopoles of charge $k$ and holomorphic rational maps 
from $S^2$ to $S^2$ of degree $k$ \cite{J}. The rational map 
may be written as $F(z)=p(z)/q(z)$,
$p(z)$ and $q(z)$ are degree $k$ polynomials in $z$ where $k$ is the monopole
charge and $z$ is a complex coordinate on the two sphere which can be
written in terms of usual polar coordinates as
$z=\tan(\theta/2)e^{i\phi}$.
The point $z$ corresponds to the unit vector 
\be
\hat{{\bf n}}_z=\frac{1}{1+|z|^2}(2\mbox{Re}(z),\,2\mbox{Im}(z),\,1-|z|^2)
\ee
The value of the rational map corresponds to the unit vector
\be
\hat{{\bf n}}_R=\frac{1}{1+|R|^2}(2\mbox{Re}(R),\,2\mbox{Im}(R),\,1-|R|^2)
\ee
Skyrmions are given by maps from $\R^3$ to $S^3$. The idea in
\cite{HMS} is to identify the domain $S^2$ of the rational map 
with concentric spheres in
$\R^3$, and the target of the rational map $S^2$ 
with spheres of latitude in $S^3$. 
A point in $\R^3$ can parametrized by $(r,\,z)\,$; $r$ denotes 
radial distance
and $z$ specifies the direction. The ansatz for the Skyrme field may
then be written as
\be
U[r,z]=\exp(if(r)\hat{{\bf n}}_R\cdot\taubf)
\ee
where $f(r)$ is a radial function satisfying $f(0)=\pi$, 
$f(\infty)=0$. $f(r)$ is determined numerically to give the closest 
approximation to the actual Skyrme configuration. 
In \cite{HMS} this ansatz was used to accurately approximate the known
minimal energy Skyrmion solutions for $B$=1 to $B$=9 and the
conjectured buckyball solution of charge 17 was shown to exist.

The Jarvis rational maps have a natural action of SO(3) given by
SU(2) M\"{o}bius transforms on the complex coordinate $z$, 
\be\label{3.4}
z\rightarrow \frac{\alpha
z+\beta}{-\bar{\beta}z+\bar{\alpha}}\,,\;\;\;\;\;\;
\;\;\;\;\;
\left(\begin {array} {cc}\alpha & \beta \\ -\bar{\beta} & \bar{\alpha}
\end {array} \right)\in\mbox{SU}(2).
\ee
This corresponds to a rotation of $\hat{{\bf n}}_z$ and generates 
rotations of the Skyrme field. For example, $z\rightarrow e^{i\theta}z$
is an anticlockwise rotation by $\theta$ 
about the $x_3$-axis, and $z\rightarrow 1/z$ is
a $\pi$ rotation about the $x_1$-axis.
An SU(2) M\"{o}bius transform on the target $S^2$ of the
rational map corresponds to a rotation of $\hat{{\bf n}}_R$, and thus to an
global isospin rotation of the Skyrme field. A rational map $F(z)$ is
symmetric under a subgroup $H$ of SO(3) if there is a set 
of pairs $\{h,\,h'\}$ such that $F(z)$ satisfies
\be\label{3.5}
F(hz)=h'F(z),
\ee
$h$ and $h'$ are SU(2) matrices with $D(h)\in H$ and
$h'$ acts on $F$ in the same manner as (\ref{3.4}). 
Note that $-h$ has the same action as $h$ in (\ref{3.5}) and similarly
for $h'$ so these are SO(3) actions.
$D(h')$ forms a representation, $\Gamma$, of $H$.
Here we are concerned only with what the symmetry group, $H$, and the
representation $\Gamma$ are, we are not at this point discussing 
the contractibility of any loops. So the double cover of $H$ does not 
enter here. Given a rational map it is easy to
determine its symmetries $H$ and the representation $\Gamma$.
The rational map ansatz accurately models the known minimal energy Skyrmion
configurations and clearly shows how the symmetry of the Skyrmion is
realised, i.e. what combination of rotations and isorotations leave
the solution invariant. 

As explained in the introduction the rational map approach is also useful
in determining the FR constraints in cases where the Skyrme
configuration needs to be split up into well separated
configurations. We will assume that whenever a monopole configuration
can be split up, respecting some symmetry then the
same can be done for Skyrmions. Generally we begin with the minimal energy
polyhedral shaped solution and end with some configuration of well
separated Skyrmions both having some discrete symmetry. From the
correspondence between monopole and Skyrmion vibrations \cite{BBT} we can see
that the Skyrmion can be vibrated keeping the relevent symmetry 
group element. The configurations are now separated 
maintaining invariance of the relevent symmetry
group element until they are far apart.
What is important is how the relative isospin
orientation of the final configurations are aligned. This is
determined by the initial vibration. In the cases we consider the
vibration corresponds to a monopole motion and thus the vibration is
of low frequency, so the Skyrmions separate in an attractive
channel. Thus the asymptotic isospin orientations are aligned to give
an attractive configuration. This will be unambiguous in 
the cases we consider. 

To determine whether a monopole configuration can be separated while 
keeping a certain symmetry, again, it is easiest to use the rational 
map description of monopoles. 
The previously described Jarvis rational maps are
suited to the description of monopoles which
are symmetric under some subgroup of SO(3). But there is no natural
action on these rational maps which corresponds to translations of the
monopoles in space. There is an equivalent rational map description 
of monopoles due to Donaldson which allows one to see how the 
monopole configuration can be separated.

In \cite{J}, the Jarvis rational map is defined by
considering solutions to the scattering equation for monopoles 
\be\label{3.6}
(D_i-i\Phi)v=0
\ee
along all radial lines through some point in $\R^3$. $D_i$ is the 
covariant derivative, $\Phi$ is the Higgs field and 
v is a complex doublet in
the fundamental representation of SU(2). The rational maps have a
natural action of SO(3) given by (\ref{3.4}) but not a simple action 
of translations, since
the choice of a point in $\R^3$ used to define the map breaks
translational symmetry. There also exists the Donaldson rational map 
which is defined by solutions to the scattering equation 
(\ref{3.6}) along all lines 
in $\R^3$ that point in a particular direction \cite{D} (which we take
here to be the $x_3$-axis). 
Donaldson rational maps for charge $k$ monopoles are
defined as based maps from $\C\rightarrow \C\cup\infty$, i.e. 
$F(w)=p(w)/q(w)$, $w\in \C$ with $q(w)$ a monic polynomial of 
degree $k$ and $p(w)$ a polynomial of degree less than $k$ with 
no factors in common with $q(w)$. Here $w$ represents a point in the
$(x_1,x_2)$-plane. 
The choice of such a direction breaks
rotational symmetry and in general there is no simple action which generates
rotations on the rational maps. But rotations about the preferred 
axis used to define the map have a simple action since they preserve this
axis. This will be enough for our purposes. Also, it is easy to 
see how translations act 
on the monopole. A rotation of angle $\theta$ about 
the preferred axis ($x_3$)
and a translation in space,
$(v_1,\,v_2,\,x)$, acts on the map as follows
\be\label{3.7}
F(w)\rightarrow e^{2x}e^{-2ik\theta}F(e^{-i\theta}(w-v))
\ee
where $v=v_1+iv_2$ and $k$ is the monopole charge.
The Jarvis rational maps are
obviously suited to the construction of monopoles and Skyrmions which
are symmetric under some subgroup of SO(3). But to see how monopoles
can be separated in space the Donaldson maps are better suited. The 
two approaches are completely equivalent as descriptions of the
monopole moduli space (we will use the notation $F(z)$ for
Jarvis maps and $F(w)$ for Donaldson maps). 

In the next section the FR
constraints will be worked out using these methods.
Once this is done it is easy to find the allowed states using (\ref{2.14}) 
and (\ref{2.15}). This determines the spin and isospin of the states.
To determine the parities of these states it is necessary to know how
the classical solution behaves under P. For Jarvis rational maps, inversion
corresponds to $z\rightarrow -1/\bar{z}$ ($\bar{z}$ 
denotes complex conjugate of $z$). 
If the rational map has a
reflection symmetry, then on the restricted configuration space of
rotations and isorotations P may be represented by some combination of
body fixed rotations and isorotations. So P may be represented by some
body fixed operator which can easily be evaluated on the angular
momentum eigenstates to give the parity eigenvalue.

\section{The $B=4$ to $B=9$ and $B=17$ Ground States}
\news
\indent
\noindent{{\bf $B=4$}}
\\[5pt]
The minimal energy $B=4$ solution has octahedral symmetry
\cite{BCT}. The octahedral group, $O_h$, is
generated by three elements, a $2\pi/3$ rotation 
cyclically permutating the Cartesian axes and a $\pi/2$ 
rotation about the $x_3$-axis, and
also the inversion element.
In \cite{LLMM}, using the instanton ansatz, it was determined  
how the octahedral
symmetry is realised so we do not need to use the rational map in this
case (the rational map approach
gives the same result). The SU(2) Skyrme field may be written as
$U_4[{\bf x}]=\sigma+i\pi^i\tau^i$ and the cubic symmetry is realised as
follows,
\bea\label{4.1}
&C_4&:\;\;\;(\pi^1,\pi^2,\pi^3)(-y,x,z)=(-\pi^2,-\pi^1,-\pi^3)(x,y,z)\\
\nonumber
&C_3&:\;\;\;(\pi^1,\pi^2,\pi^3)(y,z,x)=(\pi^2,\pi^3,\pi^1)(x,y,z)\\ \nonumber
&\mbox{Inv}&:\;\;\;(\pi^1,\pi^2,\pi^3)(-x,-y,-z)=(\tilde{\pi}^1,\tilde
{\pi}^2,\tilde{\pi}^3)(x,y,z)
\eea
where $\tilde{\pi}^1=\frac{1}{3}(\pi^1-2\pi^2-2\pi^3)$ 
and cyclically permutating.
Next, we need to work out the FR constraints associated with the $C_3$
and $C_4$ elements (the inversion element cannot be represented as a
closed loop in the configuration space and there is no FR constraint
associated with it). Since we are in the even nucleon number sector a $2\pi$
rotation in space or in isospace is a closed contractible loop and is
associated with a phase of (+1). 

The FR constraint for the
$C_3$ element is easy to determine. The $C_3$ element implies that a
$2\pi/3$ rotation combined with a $-2\pi/3$ isorotation leaves 
$U_4[{\bf x}]$ invariant. 
Simply repeat the action three
times to get a $2\pi$ rotation and a $-2\pi$ isorotation.
$2\pi$ rotations or isorotations are contractible.
Since the $C_3$ action repeated three
times is contractible this implies the $C_3$ element itself 
must be contractible so all permissable states must be eigenstates of 
the $C_3$ operator with eigenvalue (+1).
In the orientation of the Skyrme field given above the
contractibility of the $C_4$ element is not obvious. It is
helpful to do a global isospin transformation which makes this
more transparent. 

If $U[{\bf x}]=h'U[D(h){\bf x}]h'^{\dagger}$ 
then a global isospin
transformed field $\tilde{U}[{\bf x}]=AU[{\bf x}]A^{\dagger}$ satisfies 
$\tilde{U}[{\bf x}]=\tilde{h}'\tilde{U}[D(h){\bf x}]
\tilde{h}^{'\dagger}$ with $\tilde{h}'=Ah'A^{\dagger}$. 
In the orientation of (\ref{4.1}) the $\pi/2$
rotation in space about the $x_3$-axis is accompanied by a $\pi$ rotation
in isospace about the ($x_1$-$x_2$)-axis. 
We choose $A$ so that for the $C_4$ element above, the $\pi/2$
rotation in space about the $x_3$-axis is accompanied by a $\pi$ rotation
in isospace about the $x_3$-axis, (as an SO(3) rotation $D(A)$ maps the
($x_1$-$x_2$)-axis to the $x_3$-axis). By a simple homotopy argument 
it is clear that a constant
isospin transformation at every point on the closed loop will not
affect its (non)contractibility.
To show the contractibility of the $C_4$ loop we can continuously
deform the loop into one in which is obviously contractible.
Since the contractibility of a loop is invariant under homotopy, this will
show that the original loop is contractible.
The charge four cube can be deformed into two well separated 
charge two doughnuts along the $x_3$-axis. It is known from the
vibrational spectra of the $B=4$ Skyrmion \cite{BBT} that it is possible
to do this while keeping the $C_4$ symmetry. The dipole moments of the
two $B=2$ doughnuts will point in opposite directions so they attract. 
This may be seen schematically in Figure 1 (for accurate pictures
of the $B=4$ to $B=9$ and $B=17$ solutions see \cite{BS} or \cite{HMS}). 

\begin{figure}[ht]
\begin{center}
\leavevmode
\epsfxsize=11cm\epsffile{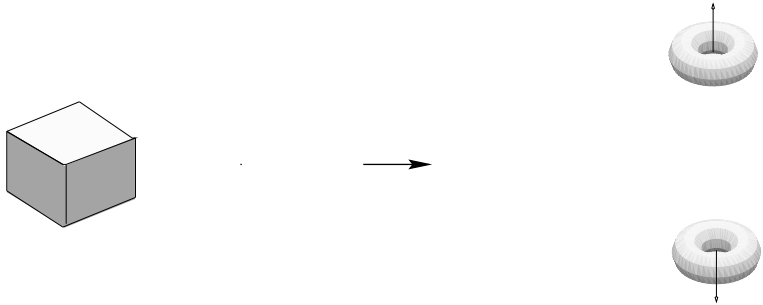}
\caption{$B=4$ Skyrmion separating to two $B=2$ Skyrmions}

\end{center}
\end{figure}

A similar type of scattering
process also occurs for monopoles and the $C_4$ symmetry is respected 
at all separations of the two 2-monopole clusters \cite{HMM}. 
The two doughnuts are
positioned at $(0,0,s)$ and $(0,0,-s)$ with $s\rightarrow 
\infty$, and are  denoted $M1$ and $M2$ respectively. 
The field may be expressed as
\be
U_4[{\bf x}]=U_2[{\bf x}-s{\bf e}_3]AU_2[{\bf x}+s{\bf e}_3]A^{\dagger}
\ee 
with $A=i(\cos\phi\,\tau_1+\sin\phi\,\tau_2)$ for some $\phi\leq 2\pi$.
$U_2[{\bf x}]$ is the axially symmetric (about $x_3$) 
charge two solution and
${\bf e}_i$ is a unit vector along the $i$ axis in space.
The form of $A$ implies that the dipole moments of $M1$ and $M2$ are
in opposite directions.
The $C_4$ symmetry implies that a simultaneous $\pi/2$ rotation 
about the $x_3$-axis with a $\pi$ isorotation about the 
$x_3$-axis leaves the configuration unchanged, i.e.  
\be
U_4[{\bf x}]\rightarrow e^{2i\lambda\frac{\tau_3}{2}}
U_4[D(e^{-i\lambda\frac{\tau_3}{2}}){\bf x}] 
e^{-2i\lambda\frac{\tau_3}{2}}
\ee
with $0\leq\lambda\leq\pi/2$.
Because it is axially symmetric $U_2[{\bf x}]$ satisfies \cite{BC}, 
\be
U_2[D(e^{-i\lambda\frac{\tau_3}{2}}){\bf x}]=e^{-2i\lambda\frac{\tau_3}{2}}
U_2[{\bf x}] 
e^{2i\lambda\frac{\tau_3}{2}}
\ee
for all values of $\lambda$.
Thus the effect of the $C_4$ transformation is a $2\pi$ isospin
transformation about the $x_3$-axis on $M2$ while leaving $M1$ unchanged.
\be
U_4[{\bf x}]\rightarrow U_2[{\bf x}-s{\bf e}_3]e^{4i\lambda\frac{\tau_3}{2}}  
AU_2[{\bf x}+s{\bf e}_3]A^{\dagger}e^{-4i\lambda\frac{\tau_3}{2}}
\ee
using $Ae^{i\lambda\frac{\tau_3}{2}}=e^{-i\lambda\frac{\tau_3}{2}}A$.
Since the $B=2$ doughnut is a boson a
$2\pi$ isospin transformation is contractible and thus the $C_4$ action
on the cube is a contractible loop and so a (+1) phase is
associated to the operator representing the $C_4$ element.
For the above  argument to work it is crucial 
that the $C_4$ symmetry is respected at
all times as the configuration is separated.

So the allowed states, $|\Psi>$ are of the form $|J,L_3>\otimes\,|I,K_3>$,
with the constraints
\bea\label{2.4.6}
e^{\frac{2\pi i}{3\sqrt{3}}(L_1+L_2+L_3)}
e^{\frac{2\pi i}{3\sqrt{3}}(K_1+K_2+K_3)}|\Psi>&=&|\Psi>\\ \nonumber
e^{i\frac{\pi}{2}L_3}e^{i\frac{\pi}{\sqrt{2}}(K_1-K_2)}|\Psi>&=&|\Psi>.
\eea
reverting to the generators used in (\ref{4.1}). 
To find the allowed states is just a matter of finding simultaneous
eigenvalues of the operators in (\ref{2.4.6}).
The ground state is
given by $|\Psi>=|0,0>\otimes\;|0,0>$; the first 
excited state with $I$=0 has $J$=4 and is 
\be
|\Psi>=(|4,4>+\sqrt{\frac{14}{5}}|4,0>+|4,-4>)\otimes\;|0,0>.\nonumber 
\ee
If $I=1$, the lowest
state has $J=2$ and is given by 
\bea
|\Psi>&=&\sqrt{6}|2,0>\otimes\,\left\{(i-1)|1,-1>+(i+1)|1,1>
\right\}\\ \nonumber
&+&\left\{|2,2>+|2,-2>\right\}\otimes\left\{2\sqrt{2}|1,0>
+(1-i)|1,1>-(1+i)|1,-1>\right\}
\eea
To compute the parities of these states we know that from the Inv
transform $U^{\dagger}(-{\bf x})=WU({\bf x})W^{\dagger}$, 
where $W=e^{\frac{i\pi}{\sqrt{3}}(K_1+K_2+K_3)}$. The parity operator
P is defined as P: $U({\bf x})\rightarrow U^{\dagger}(-{\bf x})$. So,
on the configuration space of zero modes the parity operator P can be
represented by $e^{\frac{i\pi}{\sqrt3}(K_1+K_2+K_3)}$. We may
act with P on the physical states to determine their parity. The
$I=J=0$ and $I=0,\,J=4$ states
both have (+1) parity, and the $I=1$, $J=2$ state has (-1)
parity. Thus we
find that the ground state for $B=4$ has spin and isospin zero and
positive parity in agreement with the ground state $^4_2$He$^+$.
The negative parity state with $I=1$, $J=2$ is observed as the lowest
isospin triplet state ($^4_1$H$^-$, $^4_2$He$^-$, $^4_3$Li$^-$)
\cite{N}. 
From nuclear tables there are a large number of states with $I=0$ that have
energies less than the $J=4$ state. Our scheme for quantization is
obviously very restrictive, the configuration is not allowed to vibrate
in any fashion. Including the vibrational modes and allowing the
Skyrmions to separate accounts for some of the missing
states. This we will do in section 5.
\\[5pt]
\noindent{{\bf $B=6$}}
\\[5pt]
The minimal energy $B=6$ Skyrmion has $D_{4d}$ symmetry. It can be
described in terms of a Jarvis rational map given by \cite{HMS}
\be\label{4.9}
F(z)=\frac{z^4+a}{z^2(az^4+1)},\;\;\;a=0.16i\,.
\ee
The $D_{4}$ subgroup is generated by two elements, a $\pi$ rotation 
about the $x_1$-axis and a $\pi$ rotation about the ($x_1$+$x_2$)-axis
(combining these two elements gives a $C_4$ rotation 
about the $x_3$ axis).
The elements act on the rational map by 
$F(1/z)=1/F(z)$ and $F(-i/z)=-1/F(z)$, i.e. a $\pi$ rotation
in space about the $x_1$-axis combined with a $\pi$ isorotation about the
$x_1$-axis leaves the solution invariant; and a $\pi$
rotation about the ($x_1$+$x_2$)-axis combined 
with a $\pi$ isorotation about the
$x_2$-axis leaves the solution invariant.
A closed loop corresponding to the first symmetry group element is 
\be\label{4.10}
U_6[{\bf x}]\rightarrow e^{i\lambda\frac{\tau_1}{2}}
U_6[D(e^{-i\lambda\frac{\tau_1}{2}}){\bf x}] 
e^{-i\lambda\frac{\tau_1}{2}}
\ee
with $0\leq\lambda\leq\pi$.

To determine how the FR constraints act we need to know whether the
closed loops generated by the $C_2$ elements are
contractible or not. To see that the loop in (\ref{4.10})
is noncontractible is not obvious by looking at the
polyhedral solution. It is helpful to 
continuously deform the minimal energy solution into three well
separated charge two doughnuts, one at the origin and the other two 
equidistant along the $x_3$-axis with their separation $2s$ very
large. We now show that it is possible to do this for
monopoles keeping the $C_2$ symmetry about the $x_1$-axis at all times, 
therefore by our earlier assumption the same can be done for
Skyrmions. It is easiest to see this using Donaldson rational
maps with $x_3$ as the preferred direction. Rotations 
about the $x_3$-axis have a
simple action on the rational map, given by (\ref{3.7}).  
Also, reflections can be defined on the maps \cite{HMM},
so a $\pi$ rotation about the $x_1$-axis can be defined by combining a
reflection in the ($x_1,\,x_3$)-plane and a reflection 
in the ($x_1,\,x_2$)-plane.
A rational map of degree $k$, $F(w)=p(w)/q(w)$ has $\pi$ 
rotational symmetry about the $x_1$-axis if 
\be\label{x1symm}
\frac{p(w)}{q(w)}=\frac{\overline{I(p(\overline{w}))}}
{\overline{q(\overline{w})}}\,.
\ee
Here $I(p)$ is the unique polynomial
of degree less than $k$ that satisfies $I(p)p=1\,\mbox{mod}\,q$.   
Since we are determining the contractibility of the $C_2$ rotation
about the $x_1$-axis we only need the configuration to remain
invariant under this $C_2$ element. But in fact we can separate the
configuration keeping all of the $D_{4}$ symmetry and it is 
convienient if we do this.
The most general charge six monopole with $D_{4d}$ symmetry 
is given by the Donaldson map
\be
F(w)=\frac{i\,tw^4+1}{w^6}\;,\;\;\;\;t\in\R\,.
\ee
Some value of $t$ corresponds to the minimal energy Skyrmion. Now let
$t=e^{2s}\rightarrow\,\infty$; $F(w)$ is given by
$i\,e^{2s}/w^2+1/w^6$.
Using the formula given in \cite{HSSS}, this corresponds to three
charge two monopoles lying on the $x_3$-axis, one at the origin and the
other two at $(0,0\pm\,s)$. The charge 
two monopoles must approach
axially symmetric monopoles as $s\rightarrow \infty$ since the overall
configuration has $C_{4}$ symmetry about the $x_3$-axis. 
By our previous arguments we assume that the $B=6$ Skyrmion can be split
up in the manner keeping $D_{4}$ symmetry. This is shown
schematically below.

\begin{figure}[ht]
\begin{center}
\leavevmode
\epsfxsize=9cm\epsffile{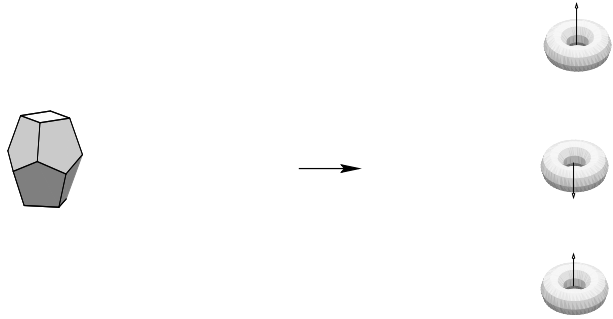}
\caption{$B=6$ Skyrmion separating to three $B=2$ Skyrmions}
\end{center}
\end{figure}

The dipole moments of the $B=2$ Skyrmion at $(0,\,0,\,s)$ and $(0,\,0,\,-s)$
point in the same direction and opposite to that of the $B=2$ Skyrmion
at the origin, so the configuration is attracting. If 
$U_6[{\bf x}]$ is of the
following form it will be $D_4$ symmetric as $s\rightarrow\infty$,
\be
U_6[{\bf x}]=U_2[{\bf x}-s{\bf e}_3]\tau_1U_2[{\bf x}]\tau_1
U_2[{\bf x}+s{\bf e}_3]\,.
\ee
Again $U_2[{\bf x}]$ is axially symmetric about the $x_3$ axis. 
Acting with the $C_2$ element has the effect of 
rotating and isorotating each of the charge two doughnuts about an axis 
in the plane of the doughnuts and also
exchanging the Skyrmion at $(0,\,0,\,s)$ with the one at
$(0,\,0,\,-s)$, this is 
\be
U_6[{\bf x}]\rightarrow U_2^a[{\bf x}]U_2^b[{\bf x}]U_2^c[{\bf x}]
\ee
where
\bea
U_2^a[{\bf x}]&=&e^{i\lambda\frac{\tau_1}{2}}U_2[D(e^{-i\lambda
\frac{\tau_1}{2}})({\bf x}-{\bf s}(\lambda))]e^{-i\lambda\frac{\tau_1}{2}}\\
U_2^b[{\bf x}]&=&
e^{i\lambda\frac{\tau_1}{2}}\tau_1U_2[D(e^{-i\lambda\frac
{\tau_1}{2}}){\bf x}]\tau_1e^{-i\lambda\frac{\tau_1}{2}}\nonumber\\
U_2^c[{\bf x}]&=&
e^{i\lambda\frac{\tau_1}{2}}U_2[D(e^{-i\lambda\frac{\tau_1}{2}})({\bf x}+
{\bf s}(\lambda))]e^{-i\lambda\frac{\tau_1}{2}}\nonumber
\eea
${\bf s}(\lambda)=sD(e^{i\lambda\frac{\tau_1}{2}}){\bf e}_3$, 
and $0\leq\lambda\leq\pi$.
The interchange of two identical doughnuts is contractible 
since they are bosons but
rotating and isorotating each of the doughnuts about an axis in their plane is a
noncontractible loop \cite{BC} and thus doing it for three doughnuts the
total loop must be noncontractible. 
As mentioned earlier, it is the noncontractibility of the above $C_2$ 
element for the charge two torus which
ensures the ground state obtained by zero mode quantizing the $B=2$
solution gives the
correct quantum numbers of the deuteron i.e. $I$=0, $J$=1 \cite{BC}. If the
loop was contractible then the ground state obtained by zero mode
quantization would have $I$=$J$=0. 

The other $C_2$ loop may be
treated in a similar manner to see that it is also noncontractible. It is
easiest to transform the field by a global isospin transform so that
the rotation and isorotation act about the same axis, then the
analysis is identical to that above.
We thus find
\bea
e^{i\pi(L_1+K_1)}|\Psi>&=&-|\Psi>\\
e^{\frac{i\pi}{\sqrt 2}(L_1+L_2)}e^{i\pi K_2}|\Psi>&=&-|\Psi>\nonumber.
\eea
This gives the ground state as 
$|1,0>\otimes\;|0,0>$. 
The first excited state with $I=0$ is $|3,0>\otimes\;|0,0>$. The lowest
state with $I=1$ is given by $|0,0>\otimes\;|1,0>$.
To determine the parity of the states we use the reflection
symmetry of the rational map $-iF(\sqrt{i}z)=\overline{F(\overline{z})}$. 
This implies that on the zero modes the parity operator can 
be represented as $P=e^{-i\frac{\pi}{2}K_3}e^{i\frac{\pi}{4}L_3}$. 
However there is an ambiguity here since the parity operator can also
be represented by the above operator times any element of $D_4$, 
since this has the same effect on the classical solution,
i.e. we could also write $P$ as
$e^{-i\frac{\pi}{2}K_3}e^{i\frac{\pi}{4}L_3}e^{i\pi(L_1+K_1)}$.
But the $C_2$ elements of $D_4$ in the ($x_1,x_2$)-plane are 
noncontractible so the operators corresponding 
to them act on the states with eigenvalue (-1). So different choices
of $P$ can give different results.
The above two choices of $P$ give opposite parity eigenvalues for all
states. We see no theoretical reason to choose one above the
other. The three states found above have the correct spins of the
corresponding ground and first excited states of $^6_3$Li and the
ground state of the isospin triplet ($^6_2$He, $^6_3$Li, $^6_4$Be). If
we choose $P$ as $e^{-i\frac{\pi}{2}K_3}e^{i\frac{\pi}{4}L_3}$ then
this gives the three states each having positive parity in agreement
with experiment. So we can choose $P$ so as to give the correct
parities of the states but theoretically there is an ambiguity 
in its definition.

A similar problem happens in the odd $B$ case. 
The $B=1$ Skyrmion is spherically symmetric 
so $P$ can be represented as the identity operator, or alternatively,
as a $2\pi$ rotation. Since $B$ is odd the two choices differ on the
quantum states. Using the convention that the nucleon have positive
parity, for $B=1$ we can take $P$ to be 
the identity operator. For all odd $B$, 
$2\pi$ rotations are noncontractible so again there are two choices of
$P$ acting on the states, $P_0$ and 
$e^{2\pi i\hat{{\bf n}}\cdot{\bf L}}P_0$ where $P_0$ is the operator
which corresponds
classically to inversion. As in the $B=6$ situation we see no way of
deciding which choice is correct. Thus, in these cases we will make
no prediction for the parities of the states.
This ambiguity may be cured by lifting to the 
full configuration space.  
This space is doubly connected for all $B$ and in the quantum 
theory states are defined on this double cover. We need to lift 
the operator $P:U[{\bf x}]\rightarrow U^{\dagger}[-{\bf x}]$ to 
the double cover. 
For $B=1$ one
chooses a lift of $P$ to the double cover of the $B=1$ configuration
space and this should determine how $P$ should be lifted for all other
$B$. But it is not obvious to us how to do this is practice.
The role of parity in the Skyrme model has also been discussed
recently in \cite{Ba}.

To summerize, for $B=6$ the states found are in agreement with the 
lowest energy states for nucleon number six, modulo our assumption about
the parity.
The ground state has spin 1 and positive parity, $^6_3$Li$^+$. 
The first excited state has spin 3 and positive parity. The lowest
state $I=1$ triplet ($^6_2$He$^+$, $^6_3$Li$^+$, $^6_4$Be$^+$),
is observed to have spin 0 and positive parity in
agreement with that found above.
\\[5pt]
\noindent{{\bf $B=8$}}
\\[5pt]
The $B=8$ case is similar to the $B=6$ case treated above. 
The minimal energy $B=8$ Skyrmion has $D_{6d}$ symmetry. It can be
described in terms of a Jarvis rational map given by \cite{HMS}
\be
F(z)=\frac{(z^6-ia)}{z^2(iaz^6-1)},\;\;\;a=0.14.
\ee
The $D_{6}$ subgroup is generated by two elements, a $C_2$ rotation 
about the $x_1$-axis and a $C_6$ rotation about the $x_3$-axes.
These act as $F(1/z)=1/F(z)$ and $F(e^{i\pi/3}z)=e^{-2\pi i/3}F(z)$.
This means that a $\pi$ rotation
in space about the $x_1$-axis combined with a $\pi$ isorotation about the
$x_1$-axis leaves the classical solution invariant; and a $\pi/3$
rotation about the $x_3$-axis combined with a $2\pi/3$ 
isorotation about the $x_3$-axis leaves the solution invariant.
Again, for the $C_2$ loop it is necessary to 
continuously deform the minimal energy solution into three well
separated charge two doughnuts,  one of charge four at the origin and one each 
of charge two equidistant along the $x_3$-axis with their separation $2s$ very
large. Then, the charge four doughnut at the origin can be separated into
two charge two doughnuts along the $x_1$-axis. 
This process can be seen to occur for monopoles in the following way. 

The most general charge eight monopole with $D_{6d}$ 
symmetry is given by the Donaldson map
\be
F(w)=\frac{i\,tw^6+1}{w^8}\;,\;\;\;\;t\in\R\,.
\ee
Again, some
value of $t$ corresponds to the minimal energy Skyrmion. Let
$t=e^{2s}\rightarrow \infty$;  
the formula given in \cite{HSSS} implies that this corresponds to two
charge two monopoles lying on the $x_3$-axis at $(0,0\pm\,s)$, and 
a charge four monopole at the origin.
The monopoles must approach
axially symmetric monopoles as $t\rightarrow \infty$ since the overall
configuration has $C_{6}$ symmetry about the $x_3$-axis. 
Next, the charge four torus can be separated into two charge two
doughnuts well separated along the $x_1$-axis keeping the $C_2$ symmetry
about the $x_1$-axis. The charge four doughnut has a Donaldson 
rational map $F(w)=1/w^4$. This can be deformed to
$F(w)=1/(w^2-v^2)^2$ for $v\in\R$ with $C_2$ symmetry about the $x_1$-axis
preserved, c.f (\ref{x1symm}).
As $v\rightarrow\infty$ this becomes two charge two doughnuts
separated along the $x_1$-axis. Again, a similar process is possible for
the charge eight Skyrmions. This is indicated below.

\begin{figure}[ht]
\begin{center}
\leavevmode
\epsfxsize=13cm\epsffile{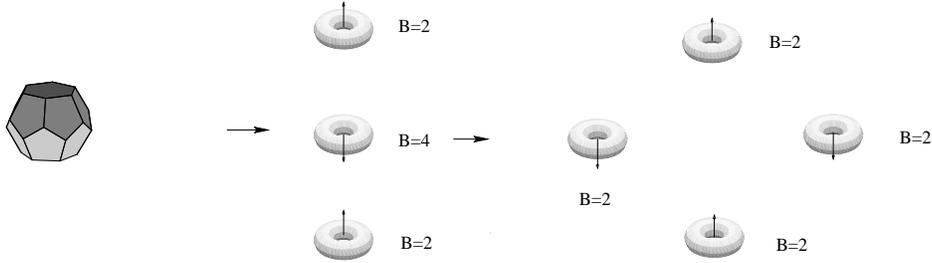}
\caption{$B=8$ Skyrmion separating to four $B=2$ Skyrmions}
\end{center}
\end{figure}

The dipole moments of the Skyrmions at $(0,\,0,\,s)$ and $(0,\,0,\,-s)$
point in the same direction and opposite to that of the charge four
Skyrmion at the origin,
so that the configuration is attracting.
Acting with the symmetry group element, which is a $\pi$ rotation and
isorotation about the $x_1$-axis,
has the effect of 
rotating and isorotating each of the four charge 2 doughnuts about an axis 
in the plane of the doughnuts and also
exchanging the Skyrmion at $(0,\,0,\,s)$ with the one at
$(0,\,0,\,-s)$. 
The interchange of two doughnuts is contractible since they are bosons;
rotating and isorotating each of the doughnuts 
about an axis in their plane is a
noncontractible loop, so doing it for four doughnuts the
total loop is contractible. The $C_6$ element can be written as a
product of the above $C_2$ element with a $C_2$ element in the 
($x_1$, $x_2$) plane at an angle $\pi/6$ to the $x_1$-axis. 
This $C_2$ loop may be seen to be
contractible in a similar manner to that above. 
So, physical states must satisfy 
\bea
e^{i\pi(L_1+K_1)}|\Psi>&=&|\Psi>\\
e^{i\frac{\pi}{3}(L_3-2K_3)}|\Psi>&=&|\Psi>\;.\nonumber
\eea
This gives the ground state as 
$|0,0>\otimes\;|0,0>$. The first excited state is given by
$|2,0>\otimes\;|0,0>$. Their parity may be determined from the reflection
symmetry of the rational map 
$e^{\frac{4\pi i}{3}}F(e^{\frac{i\pi}{6}}z)=\overline{F(\overline{z})}$. 
This implies that
on the zero modes the parity operator can be represented as 
$P=e^{\frac{i\pi}{6}(L_3-2K_3)}$. There is no parity ambiguity here 
since $B$ is even and all the FR constraints are $+1$.
Thus, both states have positive parity. 
Again, this is in agreement with the
spin 0 positive parity ground state of Beryllium 8, $^8_4$Be$^+$,
and the first excited state has spin 2 and positive parity \cite{N}. 
\\[5pt]
\noindent{{\bf $B=5$}}
\\[5pt] 
The minimal energy $B=5$ Skyrmion has $D_{2d}$ symmetry. It can be
described in terms of a Jarvis rational map given by
\be
F(z)=\frac{z(z^4-ibz^2-a)}{az^4+ibz^2-1},\;\;a=3.07,\,b=3.94.
\ee
The rational map has the symmetries $F(-z)=-F(z)$ and $F(1/z)=1/F(z)$ (for
all $a$, $b$). 
This is a simultaneous rotation and isorotation by
$\pi$ about the $x_3$-axis and a simultaneous rotation and isorotation by
$\pi$ about the $x_1$-axis.

The spin and isospin of the states 
must be half-integral since the nucleon number is odd.
In the odd nucleon sector it is necessary to be careful when
considering the (non)contractibility of the closed loops since $2\pi$
rotations are noncontractible. For instance if a configuration is
invariant under a $\pi$ rotation about some axis then the clockwise
rotation has a different FR constraint to the anticlockwise rotation
since they differ by a $2\pi$ rotation. 

To determine the (non)contractibility of the closed loops
firstly deform $F(z)$ until $b=0$. $F(z)$ now has $D_4$ symmetry
including a $C_4$ rotation and isorotation about the $x_3$-axis. 
Explicitly, $F(z)=-iF(iz)$. Or the path
\be\label{2421}
z\rightarrow e^{i\lambda}z\;,\qquad
F\rightarrow e^{-i\lambda}F\qquad 0\leq\lambda\leq\pi/2
\ee 
is a closed loop on the configuration corresponding to $b=0$. 
This path corresponds to an anti-clockwise rotation by $\pi/2$ about
the $x_3$-axis combined with a clockwise isorotation by $\pi/2$ about
the $x_3$-axis. The path traversed twice is a contractible loop since 
it is the product of two closed loops. This loop 
must also be contractible for the minimal energy solution. The path 
is now given by (\ref{2421}) with $0\leq\lambda\leq\pi$.
So an anti-clockwise rotation combined with a clockwise isorotation 
by $\pi$ about the $x_3$-axis 
is a contractible loop. From (\ref{2.141}) and (\ref{2.14}), this 
implies that the operator 
$e^{i\pi(L_3+K_3)}$ acting on the allowed states gives (+1).

Just to be clear about this suppose instead that for the solution with
$b=0$, we rotated it by $\pi/2$ anticlockwise and isorotated it by 
$3\pi/2$ anti-clockwise, again this is a closed loop, i.e.
\be\label{24211}
z\rightarrow e^{i\lambda}z\;,\qquad
F\rightarrow e^{3i\lambda}F\qquad 0\leq\lambda\leq\pi/2\;.
\ee
Repeated this loop twice gives a contractible loop which can be written
as
\be\label{242111}
z\rightarrow e^{i\lambda}z\;,\qquad
F\rightarrow e^{-i\lambda}e^{4i\lambda}F\qquad 0\leq\lambda\leq\pi\;.
\ee
This is the product of the loop in (\ref{2421}) (with  
$0\leq\lambda\leq\pi$) with a $4\pi$
isorotation which is contractible so we reach the same conclusion.
Note that the operator  $e^{i\pi(L_3-K_3)}$, which acts on states with
eigenvalue (-1), does not correspond to a closed loop traversed twice
when acting on the configuration with $b=0$ 
i.e. when $b=0$, $F(z)\neq iF(iz)$. 

Next consider the $C_2$ symmetry group element. 
It is possible to deform the minimal energy charge five Skyrmion into
a configuration of a $B=3$ tetrahedron and two $B=1$
Skyrmions on opposite sides of the tetrahedron. 
This is indicated below.

\begin{figure}[ht]
\begin{center}
\leavevmode
\epsfxsize=11cm\epsffile{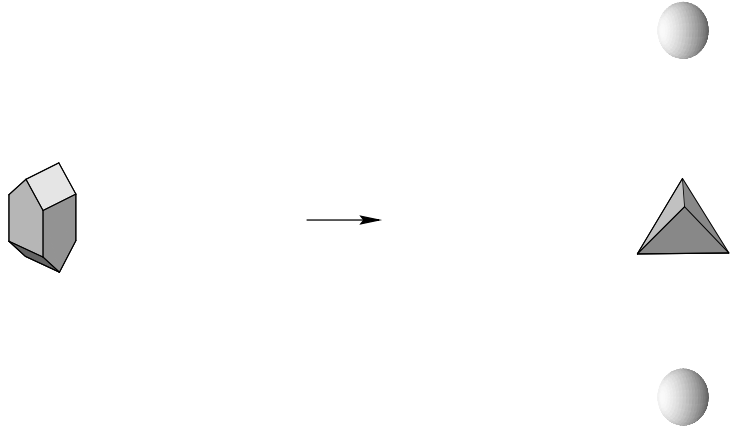}
\caption{$B=5$ Skyrmion separating to two $B$=1 Skyrmions and a $B=3$ Skyrmion}
\end{center}
\end{figure}

The $B=3$ looks like an 
anti-Skyrmion at large distances from its centre so the total
configuration is attracting. The $B=5$ solution was originally found 
by relaxing such a configuration \cite{BS}. The $B=1$ Skyrmions will be
on the $x_3$-axis equidistant from the origin with the same 
isospin orientation. The $B=3$ tetrahedron is oriented so that its
axes of second order are the $x_1$, $x_2$ and $x_3$ axes.
It is easily seen that such a configuration of monopoles 
can be separated keeping the $C_2$ symmetry about the $x_1$-axis 
at all times since the set of $k=5$ monopoles with $C_2$ symmetry about
the $x_1$-axis is connected. We take the $C_2$ element to act by an
anticlockwise rotation combined with a clockwise isorotation. The effect
of this is to rotate anti-clockwise 
and isorotate clockwise the tetrahedron and the 
$B=1$ Skyrmions by 
$\pi$ about the $x_1$-axis and interchange the two
$B=1$ Skyrmions. 
The zero mode analysis for the $B=3$ tetrahedron was considered in
\cite{C}. The loop corresponding to the 
$\pi$ anti-clockwise rotation and $\pi$ clockwise
isorotation turns out to be contractible. We will review this for the $B=9$
case which has tetrahedral symmetry, the analysis is the same as
for the $B=3$ case.
For the $B=1$ Skyrmions a rotation combined with the opposite
isorotation about the same axis leaves the configuration unchanged 
due to their hedgehog nature. The 
interchange of two identical 
$B=1$ Skyrmions is a noncontractible loop. 
Thus the overall loop is noncontractible.
So, physical states satisfy
\bea \label{4.23}
e^{i\pi(L_3+K_3)}|\Psi>&=&|\Psi>\\ \nonumber
e^{i\pi(L_1+K_1)}|\Psi>&=&-|\Psi>\,.
\eea
Since spin and isospin act in the same way we can rewrite
(\ref{4.23}) as
\bea\label{4.24} 
e^{i\pi M_3}|\Psi>&=&|\Psi>\\ \nonumber
e^{i\pi M_1}|\Psi>&=&-|\Psi>\,
\eea
where $M_i=L_i+K_i$. 
The ground state is $|M,M_3>=|1,0>$. In terms of $I$, $J$ this is 
\be
|\Psi>=|\frac{1}{2},\frac{1}{2}>\otimes\;|\frac{1}{2},
-\frac{1}{2}>-\;|\frac{1}{2},-\frac{1}{2}>\otimes\;
|\frac{1}{2},\frac{1}{2}>
\ee 
As discussed earlier we will ignore the question of parity in the odd
$B$ sector.
We recall that this is the only case where the states
$|I,K_3>\otimes\;|J,L_3>$ are not necessarily eigenstates of the
Hamiltonian since the symmetry group does not have an axis of order
higher than the second. But it is easy to see that states with
$I=\frac{1}{2}$, $J=\frac{1}{2}$ are eigenstates of the 
Hamiltonian, since the Hamiltonian only causes transitions 
from $L_3$ to $L_3+2$, $L_3$, and
$L_3-2$, and similarly for $K_3$. 
So the $I=\frac{1}{2}$, $J=\frac{1}{2}$ state is an energy eigenstate.
This is inconsistent with the observed isodoublet ground state 
($^5_2$He, $^5_3$Li)  which has spin $\frac{3}{2}$ \cite{N}.  
This state can be obtained from $|M,M_3>=|2,2>-\,|2,-2>$, which
satisfies (\ref{4.24}) but this
has higher energy than $|M,M_3>=|1,0>$.
For the Helium-Lithium isodoublet the first excited state is a 
spin $\frac{1}{2}$ state at excitation energy approximately 5 MeV. 
So the ground state we obtain is the first experimentally observed
excited state of ($^5_2$He, $^5_3$Li).

The inclusion of the vibrational modes will give new states 
but the lowest energy 
state will still be the $I=\frac{1}{2},\,J= \frac{1}{2}$ state. 
It is possible that a more careful quantization which allows the
Skyrmions to separate will raise the energy of the spin 
$\frac{1}{2}$ state above that of the spin $\frac{3}{2}$ state but 
this is not at all obvious and would be a very challenging project.
\\[5pt]
\noindent{{\bf $B=7$}}
\\[5pt] 
The minimal energy $B=7$ Skyrmion has icosahedral symmetry $Y$. It can be
described in terms of a Jarvis rational map given by
\be
F(z)=\frac{bz^6-7z^4-bz^2-1}{z(z^6+bz^4+7z^2-b)},\;\;\;b=\pm\sqrt7/5.
\ee
The icosahedral group is generated by two elements, a $C_5$ rotation 
and a $C_3$ rotation. The rotations form the defining $F_1$
representation of $Y$ (using the notation of \cite{DDH}) 
and one can check that the accompanying
isospin transformations are in the other three dimensional irreducible
representation $F_2$ (which only differs from $F_1$ in that, elements
which in $F_1$ are represented by a $2\pi/5$ rotation are represented
in $F_2$ by a $4\pi/5$ rotation).
Again we are in the odd nucleon number sector 
and so the spin and isospin of the states must be half-integral.

To determine the FR constraints is more complicated in this case.
We want to use the representation theory of the icosahedral group to
determine the allowed states. But as discussed in section 2 we
need to lift the SO(3) elements to SU(2). Generally it is not possible
to embed a group into its double group while maintaining the group
structure, i.e. to choose a subgroup isomorphic to $H$ in the group
$\bar{H}$. This means we cannot immediately 
use the representation theory of the
icosahedral group $Y$.
We need to consider the group $K$ consisting of the elements 
$\{\pm h,\,\pm h'\}$ where the elements $D(h)$ form 
the $F_1$ representation of $Y$ and the elements $D(h')$ form 
the $F_2$ representation of $Y$. The elements $h$ form the
fundamental or defining representation, denoted 
$\Gamma_6$, of the double group $\bar{Y}$. 
The elements $h'$ form the other irreducible two dimensional 
representation of $\bar{Y}$, denoted $\Gamma_7$, as can be seen 
from examining the character table of $\bar{Y}$. 
The character table of the double group $\bar{Y}$ is given above, with
$\tau=(1+\sqrt{5})/2$.
Both the representations $\Gamma_6$ and $\Gamma_7$ are
representations of the double group $\bar{Y}$ and are 
not representations of $Y$, i.e. $\Gamma_i(-y)=-\Gamma_i(y)$ 
for $i=6,\,7$, where $y$ is an abstract group element of $\bar{Y}$. 
This means that the elements in $K$ are 
of the following form 
\be\label{Keqn}
K=\left\{(\Gamma_6(y),\,\Gamma_7(y)),\;(\Gamma_6(y),\,-\Gamma_7(y)),\quad
y\in\bar{Y}\right\}\;.
\ee
\begin{table}
\caption{Character Table for $\bar{Y}$}\begin{center}
\begin{tabular}{c|ccccccccc}
&$E$&$\bar{E}$&$12C_5$&$12\bar{C_5}$&$12C^2_5$&$12\bar{C}_5^2$&
$20C_3$&$20\bar{C_3}$&$30C_2$\normalsize\\[5pt]
 \hline\\[-15pt]
  $\Gamma_1(A)$&1&1&1&1&1&1&1&1&1\\
  $\Gamma_2(F_1)$&3&3&$\tau$&$\tau$&1-$\tau$&1-$\tau$&0&0&-1\\
  $\Gamma_3(F_2)$&3&3&1-$\tau$&1-$\tau$&$\tau$&$\tau$&0&0&-1\\
  $\Gamma_4(G)$&4&4&-1&-1&-1&-1&1&1&0\\
  $\Gamma_5(H)$&5&5&0&0&0&0&-1&-1&1\\[5pt]
 \hline\\[-15pt]
  $\Gamma_6$&2&-2&$\tau$&-$\tau$&-1+$\tau$&1-$\tau$&1&-1&0\\
  $\Gamma_7$&2&-2&1-$\tau$&-1+$\tau$&-$\tau$&$\tau$&1&-1&0\\
  $\Gamma_8$&4&-4&1&-1&-1&1&-1&1&0\\ 
  $\Gamma_9$&6&-6&-1&1&1&-1&0&0&0\\ 
\end{tabular}\end{center}\end{table}
Or as a group, $K=\bar{H}\times\Z_2$. We now restrict to the group
elements $(\Gamma_6(y),\,\Gamma_7(y))$, these form a subgroup
of $K$ which is isomorphic to $\bar{Y}$,
(note that elements $(\Gamma_6(y),\,-\Gamma_7(y))$ do not form a
subgroup of $K$).
Each element in this group corresponds to a
symmetry of the $B=7$ Skyrmion and there exists a corresponding
operator which acts on the allowed states with eigenvalue $\pm 1$. 
Because $\bar{Y}$ forms a subgroup of $K$,
this implies that the states transform by a representation of the
group $\bar{Y}$. Since the states acquire only a $\pm 1$ phase under
each operation the representation must be one dimensional. 
The only such representation is the trivial one.
This means that the allowed states have eigenvalue +1 corresponding to
each of the above transforms. 
 So here we can determine the FR
constraints without any need of separating the configuration into
individual Skyrmions. This is because there are no nontrivial one dimensional
representations of the group $\bar{Y}$. In the previous cases of
$B=4,6,8$ the symmetry group of the minimal energy configuration had
$O$, $D_4$ and $D_6$ symmetries respectively. Each of these groups
have nontrivial one dimensional representations. Thus in these cases,
group theory alone cannot give the answer and it was necessary to
examine a configuration of well separated Skyrmions in order to
determine the contractibility of the loops.

Returning to the $B=7$ case, to find physical states 
of spin $J$, isospin $I$ we need to 
decompose the spin $J$ representation of SU(2) into 
representations of $\bar{Y}$ and the spin $I$ representation 
of SU(2) into representations of $\bar{Y}$. 
We then take tensor products of these
representations and look for values of $I,\,J$ that give 
singlets of $\bar{Y}$. We need to take into
account here that the isorotations are in the $F_2$ representation 
of $Y$, recall (\ref{2.141}) and (\ref{2.14}). 
We keep $I=\frac{1}{2}$, because states of high isospin are
energetically unfavourable, this means that the isospin states 
transform by the $\Gamma_7$ representation of $\bar{Y}$.
The lowest allowed $J$ is that
which its decomposition into representations of $\bar{Y}$ contains the
$\Gamma_7$ representation, since $\Gamma_7\otimes\Gamma_7$ contains the
trivial representation.
We find that the lowest $J$ is $\frac{7}{2}$ \cite{DDH}, in 
contradiction with the observed isodoublet of spin $\frac{3}{2}$.
The spin $\frac{7}{2}$ state we found appears as the
second excited state of the Lithium-Beryllium doublet at 4.6 MeV. 
The first excited state has spin $\frac{1}{2}$ at 0.5 MeV.
 
As noted earlier, it is possible to combine the vibrational modes with
the rotational modes. This will give an enlarged set of states. The
experimentally observed ground
state with $I=\frac{1}{2}$ and $J=\frac{3}{2}$ 
can be obtained in this manner. The first observed excited state with spin
$\frac{1}{2}$ can also be obtained.
Since the vibrational frequencies are as yet unknown it is not
clear whether in our analysis these states will have lower energy than the 
$I=\frac{1}{2},\,J=\frac{7}{2}$ state.
The ground state may be written as
\bea
|\Psi>&=&\{\sqrt{\frac{7}{10}}|\frac{7}{2},-\frac{3}{2}>-\sqrt{\frac{3}
{10}}|\frac{7}{2},\frac{7}{2}>\}\otimes
\;|\frac{1}{2},-\frac{1}{2}>\\ \nonumber
&+&\{\sqrt{\frac{7}{10}}{|\frac{7}{2},\frac{3}{2}>+
\sqrt{\frac{3}{10}}|\frac{7}{2},-\frac{7}{2}>\}}
\otimes\;|\frac{1}{2},\frac{1}{2}>\,.
\eea

\noindent{{\bf $B=9$}}
\\[5pt] 
The minimal energy $B=9$ Skyrmion has tetrahedral symmetry. It can be
described in terms of a rational map given in \cite{HMS}.
The rotational subgroup is generated by two elements, a $C_2$ rotation 
about the $x_3$-axis and a $C_3$ rotation about the ($x_1$+$x_2$+$x_3$)-axis. 
The rotations form the defining $F$
representation of the tetrahedral group $T$ and one can 
check that the accompanying
isospin transformations are also in the representation $F$.
Here we are in the odd nucleon number sector and again the spin and
isospin of the states must be half-integral.
To determine the FR constraints here is similar to that for the $B=7$
case. The fundamental representation of $\bar{T}$, the double group of
the tetrahedral group $T$, is denoted $\phi$. By analogy with
(\ref{Keqn}) the group $K$ is of the form 
\be\label{Keqn1}
K=\left\{(\phi(y),\,\phi(y)),\;(\phi(y),\,-\phi(y)),\quad
y\in\bar{T}\right\}\;.
\ee
So again $K$ is of the form $K=\bar{T}\times\Z_2$, and since $\bar{T}$
is a subgroup of $K=\bar{T}\times\Z_2$, states transform by a
representation of $\bar{T}$ which must be one dimensional. There are
no nontrivial representations of $\bar{T}$ \cite{LL}. 
Since the rotations act in
the same way as the isorotations the constraints can be expressed in
terms of the operators $M_i=L_i+K_i$ just as for the $B=5$ case, again
using (\ref{2.141}) and (\ref{2.14}).
Since all constraints are trivial we get
\bea 
e^{i\pi M_3}|\Psi>&=&|\Psi>\\ \nonumber
e^{i\frac{2\pi}{3\sqrt3}(M_1+M_2+M_3)}|\Psi>&=&|\Psi>\,.\
\eea 
This analysis is the same as that presented by Carson in
\cite{C} for the tetrahedrally symmetric $B=3$ solution, 
where he found the ground state to be $I=J=\frac{1}{2}$. 
The state is $|M,\,M_3>=|0,\,0>$. In terms of $I,\,J$ this is
\be
|\Psi>=|\frac{1}{2},\,\frac{1}{2}>\otimes\;|\frac{1}{2},\,-\frac{1}{2}>-
|\frac{1}{2},\,-\frac{1}{2}>\otimes\;|\frac{1}{2},\,\frac{1}{2}>\,.
\ee
Again this is not in agreement with the isodoublet of 
Beryllium and Boron of spin $\frac{3}{2}$ ($^9_4$Be, $^9_5$B). 
The state obtained is the first excited state with excitation energy
1.6 MeV \cite{N}. The observed ground state can be obtained here by
including the vibrational modes but it will have higher energy than
the spin $\frac{1}{2}$ state.
This is a similar situation to
above for $B=5$ with no obvious way around this difficulty even if the
vibrational modes are included.
\\[5pt]
\noindent{{\bf $B=17$}}
\\[5pt]
For $B\geq 9$ the minimal energy Skyrmion configurations are not yet
known. From \cite{BS} it is expected that the minimal energy solution
will look like a polygon with 12 pentagons and $2(B-7)$ hexagons. But
as $B$ increases there are many such polygons and it turns out that
the energy difference between these solutions is very small, so it is hard
to identify the minimal energy solution. But for $B=17$ a
particularly symmetric configuration arises, the buckyball solution
with icosahedral symmetry. Due to its enhanced symmetry, it is believed
that this is the minimal energy solution for $B=17$.

This solution is described by the rational map \cite{HMS}
\be\label{icos}
F(z)=\frac{17z^{15}-187z^{10}+119z^5-1}{z^2(z^{15}+119z^{10}+187z^5+17)}.
\ee
This case is similar to that for $B=7$ which also has icosahedral
symmetry. It can be checked from (\ref{icos}) that the rotations form the
defining representation $F_1$ and the isorotations form the
representation $F_2$. This
is exactly the same as for $B=7$. So we find the ground state has 
$I=\frac{1}{2}$, $J=\frac{7}{2}$. However, from
\cite{N} this state is the eighth excited state of the
isodoublet ($^{17}_8$O, $^{17}_9$F) whose ground state has spin
$\frac{5}{2}$.  

\section{Vibrational Modes}
\news

To go beyond the first approximation of just considering the zero modes  
it is appropriate to include the vibrations of the Skyrmions. These
have been calculated for the minimal energy $B=2$ and $B=4$ solutions 
\cite{BBT}.
The approximation of treating the interaction potential of
the Skyrme configurations as a harmonic oscillator potential is 
not very accurate, since, as
the minimal energy configuration separates into individual Skyrmions
the potential flattens out. A more accurate treatment will involve
estimating the inter-Skyrmion potential at intermediate and large 
separations. Thus it should not be expected 
that the inclusion of vibrational modes will yield
accurate results for masses, binding energies of states etc.
 
Including the vibrational modes involves the coupling of harmonic 
oscillator wavefunctions to the rotational and isorotational
wavefunctions. However, they
do not combine in an arbitrary way; the interaction of the
rotations and vibrations is described in \cite{GK} for general soliton
models. The space of rotations and isorotations is 
$(\mbox{SO}(3)\times\mbox{SO}(3))/H$; again $H$ is the 
symmetry group of the minimal energy
solution. The vibrations fall into representations of $H$ and the space
of vibrations is a vector space denoted by $V$. $V$ is a direct sum of
vector spaces $V_i$ with $H$ acting irreducibly on each $V_i$.

The total configuration space ${\cal F}$ say, is now a vector bundle over 
$(SO(3)\times SO(3))/H$. For ease of notation we will restrict here to
the case of even $B$ so we don't need to worry about the double
covering. It can be included without much difficulty.
${\cal F}$ can be defined by taking the product space
$\mbox{SO}(3)\times\mbox{SO}(3)\times V$ with the following equivalence
\be\label{vibs}
(R,\,R',\,v)\cong (SR,\,R'\Gamma^{-1}(S),\,\rho^{-1}(S)v)\,,\;\;S\in H,\,
\;\;(R,R')\in \mbox{SO}(3)\times\mbox{SO}(3)\;,\;\;v\in V
\ee
$\Gamma(S)$ is as before 
and $\rho(S)$ is the action of $H$ on the space of vibrations,
As an example to see that this gives the correct 
configuration space consider the
$B=4$ Skyrmion which has a cubic shape. One of the vibrational modes is
the so called tetrahedral mode which can be imagined as follows. The
vertices of the cube form two interlocking tetrahedra. The vibrating
cube alternately separates into four Skyrmions on the vertices of one of the
tetrahedra (positive mode), then contracts to the cube and 
then separates into four Skyrmions on the vertices of the dual 
tetrahedron (negative mode). Acting with the $\pi/2$ 
rotation and $\pi$ isorotation about the $x_3$-axis 
(which is a symmetry of the cube) is
equivalent to interchanging the positive and negative modes. So as not
to overcount the configuration space we must identify rotating and
isorotating the configuration about the $x_3$-axis with 
interchanging positive and negative vibrating modes.

Quantum states are given by the direct product of Wigner functions on
SU(2)$\times$ SU(2) with harmonic oscillator wave functions on $V$ with
the proviso that the states are $H$ invariant. Again, in a manner
similar to that treated for the zero modes the FR constraints 
determine how $H$ invariance is to be implemented. The FR constraints
for the closed loops corresponding to the above action of $H$ 
are identical to those when just considering zero modes. This is
because the loops are closed for all vibrational amplitudes, so the
loop can be deformed to the case of amplitude zero, i.e. the zero mode
case. When the classical solution has a reflection symmetry the
vibrations corresponding to the vector space $V_i$ have a definite 
parity $\rho_i=\pm 1$. It is possible
to check that the parity operator for the rotational and 
vibrational states is given by $P\prod_i\rho_i^n$ where $P$ is the parity
operator acting on the zero modes and the $i$th vibrational state is in the
$n$th excited mode.

Here we will concentrate on the $B=4$ case since the vibrational
spectra has been calculated \cite{BBT}. The spectra was
calculated at finite pion mass, whereas we are working with zero pion
mass. But the vibrational frequencies found in \cite{BBT} do not
appear to vary greatly with the value of the pion mass used, so we will
use their values. Anyway we are not interested in obtaining accurate numbers
here, we just want to indicate how to couple the rotational and
vibrational modes.

To find the allowed states is quite easy. If one is only
interested in what states are allowed and not their dependence in terms
of $L_3, K_3$ etc., then this can be determined by the representation
theory of the cubic symmetry group $O$, alone. The
configuration space is $\mbox{SO}(3)\times\mbox{SO}(3)\times V$ 
quotiented by $O$ as described above. 
Since all the FR constraints all $+1$, the allowed
states are $O$ singlets of SO(3)$\times$SO(3)$\times V$. From
(\ref{4.1}) we know that the rotational SO(3) transforms as the defining
$F_1$ representation of $O$ and that the isorotational SO(3)
transforms as the $E\oplus A_2$ representation of $O$,
using the notation of \cite{LL}. From this we
can work out how a spin $J$, isospin $I$ state decomposes under $O$. The
representations of $O$ that the vibrations form were computed
in \cite{BBT} and so we can determine how the product 
SO(3)$\times$SO(3)$\times V$ transforms under $O$ and so we can easily
read off which combinations of $I,\,J$, and vibrations are allowed as states.
For $B=4$ the rotational moments of inertia are all equal, 
$V_{ij}=\delta_{ij}$ (18 MeV)$^{-1}$, the isorotational
moments of inertia are $U_{11}=U_{22}=(82.2$ $\mbox{MeV})^{-1}$, 
$U_{33}=(68.2$ $\mbox{MeV})^{-1}$,
and the cross term between spin and isospin vanishes,
$W_{ij}=0$. These values are obtained using the values 
of $f_{\pi}$, $e$ from \cite{ANW}. 
Thus the Hamiltonian is 
\be
H=41.1{\bf K}^2-7.0K_3^2+9.0{\bf L}^2,  
\ee
in units of MeV. The energies of the vibrational states,
$\hbar\omega$, and their representations of the cubic group are 
($E^+$, 94 MeV), ($A^-_2$, 104 MeV), ($F_2^+$, 107 MeV), 
($F_2^-$, 132 MeV), ($A_1^+$, 155 MeV), ($F_2^-$, 168 MeV), 
and ($F_2^+$, 189 MeV), the $\pm$ denotes parity.
Restricting to $K=0$, i.e. $^4_2$He, the first few excited states are
$J=2^+$ at 147 MeV, $J=0^+$ at
155 MeV, $J=2^+$ at 160 MeV, and then the first excited zero mode
state, $J=4^+$ at 178 MeV.  
The observed excited states of $^4_2$He$^+$ differ considerably from
this \cite{N}. The first few excited states are $0^+$ at 20.1 MeV,
$0^-$ at 21.1 MeV, and $2^-$ at 22.1 MeV. The most obvious discrepancy
is the over estimation of the excitation energies, this is partly due
to treatment of the potential as of harmonic oscillator type. 
Nonetheless this shows that the vibrational states are important and
are of the same order of energy as the pure rotational states.

The experimentally observed ground state of ($^7_3$Li, $^7_4$Be) has
$J=\frac{3}{2}$. For $B=7$ the lowest state with isospin
$I=\frac{1}{2}$ was found to have $J=\frac{7}{2}$. By the 
same methods as above,
using the monopole vibrations as a prediction for the low lying
Skyrmion vibration frequencies, a state of $J=\frac{3}{2}$ can be
obtained. If the vibrational frequency of this state is not too high
it may have lower energy than the $J=\frac{7}{2}$ and thus give the
correct ground state. 

\section{Nucleon Densities of the States}
\news

Given the expressions for the states in terms of Wigner functions, other
physical properties may be calculated such as the nucleon density of
the quantum state. The nucleon density of the classical configurations 
are quite symmetrical and it is of interest to know how quantum
effects change this. As discussed in the introduction, to compare
with experiment it would be
desirable for the quantum states to be mostly or completely S-wave, 
and we shall see that this is the case.
Given a state $\Psi$, we want an expression for
the probability distribution $p_{\Psi}({\bf x})$ on physical space
which is interpreted as the nucleon density. We do this 
by averaging the classical nucleon density over the space of zero modes
weighted with $|\Psi|^2$ \cite{LMS} (we restrict here to zero mode states). 
Denoting the classical nucleon density by $B({\bf x})$, the spatial
probability distribution for the quantum state is defined as 
\be\label{6.1}
p_{\Psi}({\bf x})=\frac{1}{2}\int B(D(A){\bf x})|\Psi(A,\,A^{'})|^2
\sin \tilde{\theta}\,d\tilde{\theta}\,d\tilde{\phi}\,d\tilde{\psi}
\ee
where $D(A)$ is parametrized by the Euler angles 
$(\tilde{\theta},\,\tilde{\phi},\,\tilde{\psi})$.
$p_{\Psi}({\bf x})$ is evaluated by expanding $B(D(A){\bf x})$ 
in terms of spherical harmonics $Y_{mn}(\hat{\tilde{{\bf x}}})$, with 
$\tilde{{\bf x}}=D(A){\bf x}$, then using
the transformation properties of spherical harmonics under rotations
\be
Y_{lm}(\hat{\tilde{{\bf x}}})=\sum_kD^{l}_{mk}(A)^*Y_{lk}(\hat{\bf
x})\;,\;\;\;\;\;(\mbox{no sum on } l)
\ee
and the fact that $|\Psi|^2$ can be written as a sum of terms 
$D^J_{ab}(A)D^J_{cb}(A)^*$, where $D^J_{ab}(A)$ are Wigner functions.
The direct product and orthogonality
properties of the Wigner functions are then 
used to compute $p_{\Psi}({\bf x})$.
We choose the space fixed angular momentum in the $x_3$-direction,
$b$, equal to $J$, i.e. ``spin up''.
If we only consider rotational and isorotational wave
functions $p_{\Psi}({\bf x})$ will have the same radial dependence 
as the classical solution. But the angular dependence will be changed
by quantum effects. In the Skyrme model there is no decomposition 
of angular momentum into orbital and spin angular momentum. However, 
calculating the spatial probability distribution can give 
some insight into what the spin and
orbital contributions of the nuclear state are. If the spatial 
probability distribution is almost spherical then it is reasonable to 
deduce that the orbital angular momentum is almost all S-wave.   
For all the examples treated below the quantum nucleon density is 
more spherically symmetric than the classical nucleon density, 
it being exactly S-wave in a number of cases, in these 
cases we conclude that all the angular momentum
is due to the spin of the nucleons. 

For $B$=4 we found the ground state to have $I=J=0$, the first excited
state with $I=0$ has $J=4$ and the lowest state with $I=1$ has $J=2$. 
Inserting the above states into (\ref{6.1}) we
trivially find the probability distribution of the $I=0$, $J=0$ state to
be spherically symmetric. This is also true of the ground state for $B=8$.
For the $I=0,\,J=4$ state of $B=4$ we find the angular dependance 
to be mostly S-wave with $l=4$ contributions and some very small 
$l=6$ and $l=8$ contributions,  
\be
p_{\Psi}(\theta,\,\phi)\propto\{Y_{00}-0.045Y_{40}-0.027(Y_{44}+
Y_{4-4})+0.0002Y_{60}+0.00003Y_{80}\}\,,
\ee
here $(\theta,\,\phi)$ are the angular coordinates on physical 
space, as opposed to the coordinates on $D(A)$.
And for the $I=1,\,J=2$ state we again find the 
nucleon density to be mostly
spherically symmetric with a small $l=4$ contribution.  
\be
p_{\Psi}(\theta,\,\psi)\propto\{Y_{00}-0.01Y_{40}-0.01(Y_{44}+Y_{4-4})\}\,.
\ee 
Thus when quantum effects are included the nucleon density becomes
spherical or near spherical. It is known that the ground state of 
$^4_2\mbox{He}$ is completely S-wave. In real nuclei the nucleon
density is large up to a certain radius and then falls off
quickly. Our quantum states have the same radial dependence as the
classical solutions which is somewhat hollow, this becomes very
noticeable for larger nucleon numbers.     

For the $I=0$, $J=1$ ground state of $B=6$ it is found that 
\be
p_{\Psi}(\theta,\,\phi)\propto\{Y_{00}-0.03Y_{20}\}\,. 
\ee
This result is slightly different than for the $B=2$ deuteron. In both
cases the ground state is given by $I=0$ and $J=1$ with the same $L_3$
dependence but for the deuteron
the quantum probability distribution is of a dumbell shape
\cite{LMS}. Here, for the $B=6$ solution the quantum probability 
distribution is of a toroidal shape. The difference arises because the
classical nucleon densities of the two solutions are different.
Nonetheless, the wave function 
is predominately S-wave and this is also in agreement with experiment.

The ground state for $B=7$ may be written as
\bea
|\Psi>&=&\{\sqrt{\frac{7}{10}}|\frac{7}{2},-\frac{3}{2}>-\sqrt{\frac{3}
{10}}|\frac{7}{2},\frac{7}{2}>\}
\otimes\;|\frac{1}{2},-\frac{1}{2}>\\ \nonumber
&+&\{\sqrt{\frac{7}{10}}{|\frac{7}{2},\frac{3}{2}>+
\sqrt{\frac{3}{10}}|\frac{7}{2},-\frac{7}{2}>\}}
\otimes\;|\frac{1}{2},\frac{1}{2}>\;.
\eea
From this we can see that the probability distribution of this
state must be spherically symmetric. This is so because in (\ref{6.1}) we
take $|\Psi|^2$ and integrate it with the classical nucleon density. 
The classical nucleon density has icosahedral symmetry and for
$l\leq 7$ the only spherical harmonics which are icosahedrally
symmetric are $l=0$ and an $l=6$ harmonic \cite{DDH}. But $|\Psi|^2$
expanded in terms of Wigner functions has no $l=6$ term and so
$p_{\Psi}({\bf x})$ is spherically symmetric. The same analysis
applies to the ground state of $B=17$.

For the $B=9$ ground state it is easy to show that the nucleon density is
spherically symmetric. Since the spin is $\frac{1}{2}$ the nucleon 
density could only have $l=0$ and $l=1$ components. But the 
$l=1$ component is associated
with a vector in space and this is incompatible with tetrahedral
symmetry so the wave function is completely S-wave. It can also be
checked that the $B=5$ ground state is completely S-wave. 

So we see that when one includes quantum effects the classical picture of
the nucleon density having a discrete point symmetry group is
changed so that in the quantum state it is smeared forming a spherical
or near spherically symmetric configuration.

\section{Outlook}
\news

We have described the ground states of the $B=4$ to $B=9$ and $B=17$
Skyrmions obtained by quantizing the zero modes of the classical
solutions. We did not attempt to calculate the masses, binding
energies and  other observables since a zero mode quantization 
is too restrictive to get accurate results. 
Nonetheless we expected to obtain the correct
quantum numbers of the ground states. However our results are not
promising; for $B=4$, $B=6$ and $B=8$ the correct ground states are
obtained. But in the odd nucleon sector we have obtained the incorrect
ground states. For nucleon numbers 5, 7 and 9 the experimentally
observed ground states are isodoublets with spin
$\frac{3}{2}$ and for nucleon number 17 the observed ground state is an
isodoublet with spin $\frac{5}{2}$.
However we obtained isodoublets with spin $\frac{1}{2}$
for $B=5$ and $B=9$, and an isodoublet with spin $\frac{7}{2}$ for
$B=7$ and $B=17$. The symmetry of the classical solutions 
which can give spin $\frac{3}{2}$ states is $C_4$ symmetry, 
and $C_6$ symmetry can give a spin $\frac{5}{2}$ state. 
But the classical solutions in these cases do not appear to 
have $C_4$ or $C_6$ symmetry.
 
The main assumption made was that certain
closed loops in the configuration space remain closed as the minimal energy
configuration is separated into $B=1$ or $B=2$ Skyrmions. This was
necessary in order to determine the FR constraints. The vibrational
spectra of the minimal energy $B=2$ and $B=4$ Skyrmions for low
frequencies is in correspondence with monopole vibrations about the
corresponding monopoles. We assumed that this correspondence holds 
true for higher $B$. We consider this very likely, but the vibrational
spectra for the Skyrmions needs to be found to confirm this.
It was also assumed that if the solution could be vibrated, remaining 
invariant under a certain symmetry, then the continuation of the
symmetric path in the configuration space results in a configuration 
of well separated Skyrmions. We have seen that this is true for 
monopoles in the cases considered and presumed it also holds for Skyrmions.
Again, this does not seem to be a particularly strong 
assumption. In any case, the
outcome for $B=7$, $B=9$ and $B=17$ is independent of these
assumptions, since the FR constraints can be determined from the group
theory alone, and the ground states obtained are not in 
agreement with experiment.

To obtain the experimentally observed ground states it will be
necessary to include modes whereby the
Skyrmions separate. It is not difficult to see that a quadratic
approximation (by just considering the vibrational modes) will not
cure this problem for the $B=5$ and $B=9$ cases.
If the Skyrme model is to correctly predict the ground states of these
nuclei it will be necessary to include configurations of 
Skyrmions with intermediate or long range separations which is a 
highly nontrivial problem. 

Another possible resolution is that 
that the solutions found in \cite{BS} are not
well defined minima, i.e. there may be a number of solutions with 
approximately equal energies and so an expansion about just one of
these minima is not valid. However we view that the more likely 
answer is that the zero mode configuration space is too restrictive.
The zero mode approximation allows only for a collective motion 
of the Skyrmions with nine parameters, while the space that 
approximates the low energy behaviour of $B$ Skyrmions should 
be $6B$ dimensional. As $B$ increases the validity of the
zero mode approximation should break down.
\\[5pt]
\noindent{\bf Acknowledgements}

I would like to thank Nick Manton for suggesting this problem and for
many fruitful discussions and comments. 
I also thank Kim Baskerville, Richard Batty, 
Conor Houghton, Arthur Mountain, 
Paul Sutcliffe and Neil Turok. I also thank PPARC for financial assistance.
\\[1cm]

\end{document}